\documentclass[twocolumn]{aastex63}
\usepackage{amsmath}
\usepackage{amsfonts}
\usepackage{amssymb}
\usepackage{bm}
\usepackage{color}
\usepackage{graphicx}
\usepackage{natbib}
\usepackage[utf8]{inputenc}

\shorttitle{A Refined Model of Convectively-Driven Flicker in Kepler Light Curves}
\shortauthors{Van Kooten, Anders, \& Cranmer}

\received{February 9, 2021}
\revised{April 7, 2021}
\accepted{April 12, 2021}
\submitjournal{\apj}

\newcommand{\fe}[1][]{\ensuremath{F_\text{8#1}}}
\newcommand{\feobs}{\fe[,obs]}
\newcommand{\femod}{\fe[,mod]}
\renewcommand{\max}{\ensuremath{_\text{max}{}}}
\newcommand{\RMS}{\ensuremath{_\text{RMS}{}}}
\newcommand{\eff}{\ensuremath{_\text{eff}{}}}
\newcommand{\Teff}{\ensuremath{T\eff}}
\newcommand{\logg}{\ensuremath{\log g}}
\newcommand{\feh}{[\text{Fe/H}]}
\newcommand{\Ma}{\ensuremath{\mathcal{M}}}
\newcommand{\CBP}{\ensuremath{C_\text{BP}}}
\newcommand{\kep}{\textit{Kepler}}
\renewcommand{\sun}{\ensuremath{_\odot}}

\newcommand{\muram}{MURaM}
\newcommand{\grad}{\ensuremath{\nabla}}
\renewcommand{\dot}{\ensuremath{\cdot}}

\begin{document}

\title{A Refined Model of Convectively-Driven Flicker in Kepler Light Curves}

\author[0000-0002-4472-8517]{Samuel J. Van Kooten}
\affil{Department of Astrophysical and Planetary Sciences, University of Colorado, Boulder, Colorado, USA}

\author[0000-0002-3433-4733]{Evan H. Anders}
\affil{Center for Interdisciplinary Exploration and Research in Astrophysics, Northwestern University, Evanston, Illinois, USA}

\author[0000-0002-3699-3134]{Steven R. Cranmer}
\affil{Department of Astrophysical and Planetary Sciences, University of Colorado, Boulder, Colorado, USA}

\correspondingauthor{Samuel Van Kooten}
\email{samuel.vankooten@colorado.edu}

\begin{abstract}
	Light curves produced by the \kep\ mission demonstrate stochastic brightness fluctuations (or ``flicker'') of stellar origin which contribute to the noise floor, limiting the sensitivity of exoplanet detection and characterization methods.
	In stars with surface convection, the primary driver of these variations on short (sub-eight-hour) timescales is believed to be convective granulation.
	In this work, we improve existing models of this granular flicker amplitude, or \fe, by including the effect of the \kep\ bandpass on measured flicker, by incorporating metallicity in determining convective Mach numbers, and by using scaling relations from a wider set of numerical simulations.
	To motivate and validate these changes, we use a recent database of convective flicker measurements in \kep\ stars, which allows us to more fully detail the remaining model--prediction error.
	Our model improvements reduce the typical misprediction of flicker amplitude from a factor of 2.5 to 2.
	We rule out rotation period and strong magnetic activity as possible explanations for the remaining model error, and we show that binary companions may affect convective flicker.
	We also introduce an ``envelope'' model which predicts a range of flicker amplitudes for any one star to account for some of the spread in numerical simulations, and we find that this range covers 78\% of observed stars.
	We note that the solar granular flicker amplitude is lower than most Sun-like stars.
	This improved model of convective flicker amplitude can better characterize this source of noise in exoplanet studies as well as better inform models and simulations of stellar granulation.
\end{abstract}


\section{Introduction}
\label{sec:intro}

While primarily intended for exoplanet discovery, the \kep\ mission's long-duration light curves with high photometric precision have proven extremely valuable for investigation of the mission's target stars as well.
These stellar studies are important in their own right, but they also provide a benefit back to exoplanet science, as a better understanding of the stellar flux allows the exoplanetary signal to be more precisely separated from the stellar signal and allows the uncertainty in the process to be better described.

One topic of interest for many investigators has been short-period variation (called \textit{flicker}) in the \kep\ light curves of F, G, and K stars with convective surface-layers.
\citet{Bastien2013,Bastien2016} defined the quantity \fe, the root mean square (RMS) amplitude of the flicker that occurs on $<8$~hr timescales, and showed that it displays a very strong dependence on stellar surface gravity.
The evolving pattern of granulation, the convectively-driven warm and cool regions at the photosphere that is well-known in solar observations, has been shown to be a very plausible and likely driver of \fe\ \citep{Cranmer2014}, with the gravity dependence explained by the very strong dependence of granular size- and timescales on surface gravity.
Indeed, granulation and magnetic activity taken together have been shown to be sufficient to very closely reproduce the entirety of solar photometric variability (aside from the characteristic 5~min signal of acoustic oscillations) over timescales from minutes to decades \citep{Shapiro2017}.

\begin{figure*}[tp]
	\centering
	\includegraphics[width=\textwidth]{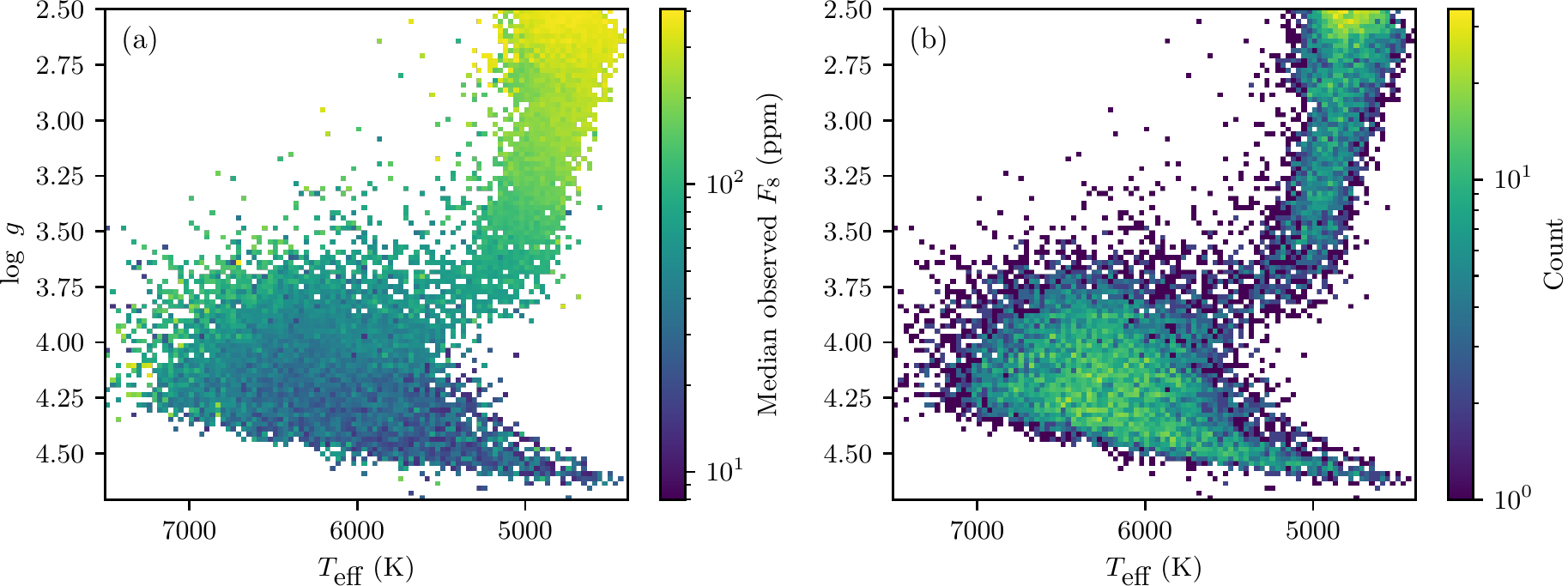}
	\caption{Our sample of 16,992 stars with both \fe\ measurements and \feh\ measurements as described in Section~\ref{sec:data}. Both panels are 2D histograms showing, in panel (a), the median observed \fe\ per bin, and in panel (b), the number of stars per bin. Note that \logg\ is strongly influenced by a star's radius and therefore luminosity, and so the \logg\ axis is effectively a skewed luminosity axis and is plotted to preserve the traditional H-R diagram orientation. The main sequence runs along the bottom edge of the data cloud, and the red giant branch ascends toward and beyond the upper-right corner.}
	\label{fig:flicker_data}
\end{figure*}

One application of \fe\ measurements or related analyses is as a photometric proxy for surface gravity \citep{Bastien2013,Bastien2016,Pande2018} or stellar density \citep{Kipping2014}, especially where spectroscopic or asteroseismic measurements may not be available.
A related approach is to measure the \textit{timescales} of flicker, rather than the amplitude, as a proxy for surface gravity \citep{Kallinger2016}.
This approach is more resilient to noise and, when used to measure surface gravity, has the advantage that flicker timescales are more directly tied to surface gravity than are flicker amplitudes.
However, the approach requires that flicker timescales be temporally resolved, which is not always the case for \kep\ long-cadence data (the data available for the vast majority of \kep\ targets), where the 30~min cadence fails to resolve, for instance, G-dwarf granular timescales on the order of 10~min.
An additional approach that has been demonstrated is to use machine learning to extract stellar parameters such as surface gravity from the power spectra of stellar variability \citep{Sayeed2020}.

Another motivation for understanding stellar flicker is to constrain the noise present in planetary transit measurements.
Studies have shown that stellar granulation produces a non-negligible effect in the noise present during observations of planetary transits, introducing, e.g., uncertainty in planetary radius measurements of a few percent \citep{Chiavassa2017,Morris2020}, or even up to 10\% \citep{Sulis2020}, depending on the observation being modeled.
One cause of this noise is that the granulation provides one of many deviations from a perfect black body, affecting signals especially during spectroscopic observations.
Another is that the granular pattern, and thus the integrated flux of the star, varies during a transit (compare transit timescales of ~hours to granular timescales of $\sim10$~min for Sun-like stars).
A third is that the transit depth varies because the planet occults only small portions of the stellar disk, which consist of a mixture of granules and intergranular lanes in a ratio that varies from one local patch to another.
From a related, stellar perspective, preliminary simulation work from \citet{Bonifacio2018} has shown that neglecting the effect of granulation can introduce errors when inferring effective temperatures from photometric colors, as much as 200~K in the most extreme cases.
When mitigating these sorts of flicker-induced uncertainties, one option is to use techniques such as Gaussian Process regressions to fit and account for the stochastic flicker signal on a star-by-star basis \citep{Pereira2019}.

Efforts have been made to model the properties of granular flicker empirically \citep[e.g.][]{Corsaro2017,Tayar2018}.
\citet{Samadi2013a,Samadi2013b} derived an analytic model for flicker, and past work has carefully compared this model with observations and sought to reconcile the differences \citep{Cranmer2014}.
In this work we improve upon this modeling effort, and we compare the model with a much larger sample of stars with measured flicker (approximately 30 times as many stars).
We describe this larger sample in Section~\ref{sec:data}.
In Section~\ref{sec:model} we present the complete \fe\ model, including our additions.
In Section~\ref{sec:results} we compare the model predictions to the observations, noting improved agreement with our model corrections.
We also pay special attention to the solar value of \fe, and we discuss the correlation, or lack thereof, of \fe\ with other observables.
In Section~\ref{sec:discussion} we discuss the remaining difference between observations and model predictions, and we finally summarize our work in Section~\ref{sec:conclusions}.

\section{Observational Data}
\label{sec:data}

For this analysis we assembled a data catalog from a variety of sources, totaling 16,992 F, G, and K stars from the \kep\ catalog, shown in Figure~\ref{fig:flicker_data}.
This improves upon the past work of \citet{Cranmer2014}, which used a smaller sample of 508 stars with very few K dwarfs.
These stars have effective temperatures \Teff\ that range from 4390 to 7500 K, with \logg\ values ranging from 2.5 to 4.7 and typical field-star values of \feh\ ranging from --1 to +0.5.

We assembled our catalog by beginning with the 27,628 \fe\ values of \citet{Bastien2016}.
Since their work focuses on using \fe\ as a proxy for \logg, they report \logg\ values derived from their \fe\ measurements.
We therefore recover the original, observed (and corrected and calibrated) \fe\ values from their published \logg\ values by inverting their (monotonic, one-to-one) \fe --\logg\ relation\footnote{We note that, while the 16-pt RMS values used to calculate \fe\ are also reported for each star by \citet{Bastien2016}, we have been informed via private communication that errors were inadvertently introduced in those values when preparing their table, with an erratum forthcoming. This motivates our more indirect route to recover the observed \fe\ values.}.
These \fe\ measurements, which measure the RMS amplitude of the portion of a star's variability that occurs on sub-eight-hour timescales, were extracted from \kep 's long-cadence, PDC-MAP light curves by subtracting from each light curve a smoothed version of itself, using an 8 hr smoothing window \citep[following][]{Basri2011,Bastien2013}.
Additional steps in their pipeline remove transient events such as flares and transits, subtract out the portion of the \fe\ measurement attributable to shot noise, and account for \kep\ pointing offsets and flux contamination from neighboring stars.

We merge this \fe\ catalog with the \Teff, \logg, and mass values of \citet{Berger2020}, which cover all but 848 of the stars with measured \fe\ values (a total of 26,780).
These values, part of a catalog intended to provide a comprehensive source of stellar parameters for \kep\ targets, are drawn from spectroscopic observations combined with \textit{Gaia} parallaxes and modeled isochrones.
The use of these \logg\ values from an independent source, rather than using those of \citet{Bastien2016}, ensures some independence between the \logg\ and \fe\ values we use.

Next, we merge into our catalog \feh\ metallicity data from the LAMOST-Kepler project \citep{Zong2018}, an effort to use the LAMOST telescope for spectroscopic follow-up observations of \kep\ targets.
This catalog covers an evenly-distributed 18,773 of the 26,780 \kep\ stars in both the \citet{Bastien2016} and \citet{Berger2020} catalogs, and we limit our analysis to this smaller sample.
While \feh\ values are reported by \citet{Berger2020}, their values are derived from multiple sources and they describe pipeline-to-pipeline variation as their dominant source of uncertainty for \feh.
We thus restrict ourselves to this one source to ensure a higher level of consistency.
We convert from \feh\ to heavy-element mass fraction $Z$ with a reference solar value of $Z\sun = 0.01696$ \citep{Grevesse1998}.

Finally, we remove all stars with $\logg < 2.5$, as the granular timescales for these large giants extend well beyond the 8-hour window that determines \fe\ \citep{Bastien2016}, and a handful of outlier stars with $\Teff > 7500$~K.
This removes 1,701 stars, yielding the final sample of 16,992 stars which is shown in Figure~\ref{fig:flicker_data}.
This final catalog is included in our code and data archive \citep{VanKooten2021_Zenodo}.

\section{Granulation Model}
\label{sec:model}

\citet{Samadi2013a,Samadi2013b} derived a theoretical model predicting the RMS amplitude $\sigma$ of granular flicker as a function of a star's effective temperature \Teff, surface gravity \logg, and mass $M$.
The model combines first-principles geometrical arguments, analytic derivations, and scaling relations, with some components further fit to numerical simulations.
Our use of this model closely follows that described by \citet{Cranmer2014}, including the conversion factor between the model-predicted $\sigma$ and the observational value \fe\ (i.e. the total RMS granular flicker amplitude versus the RMS amplitude of the flicker component occurring over $<8$ hour scales).
However, in this work we show that the model can be cast in more absolute terms, rather than as a scaling relation relative to the solar $\sigma$.
We also present updated functions for predicting: (1) the Mach number of near-surface, vertical plasma flows, (2) the relative temperature contrast between granular centers and lanes, and (3) the characteristic size of granules.
We also add a correction factor for the influence of \kep 's bandpass on observed \fe\ values.

Here we present in full the version of the model used in this work.
Our Python code implementing this model and producing our plots is included in our code and data archive \citep{VanKooten2021_Zenodo}.

\newpage

\subsection{Bolometric Flicker}
\label{sec:bolo-amplitude}

\citet{Samadi2013a} derive an expression for $\sigma_\tau$, the RMS amplitude of the bolometric intensity variation in a stellar light curve due to granulation seen at optical depth $\tau$, of
\begin{equation}
	\sigma_\tau = \frac{12}{\sqrt{2}} \sqrt{\frac{\tau_g}{N_g}}\;\Theta\RMS^2,
	\label{eqn:sigma_tau}
\end{equation}
where $\tau_g$ is the characteristic optical thickness of granules (which is very nearly constant across the F, G and K dwarfs and giants used in this work), $N_g$ the average number of granules covering the visible half of the star, $\Theta\RMS$ is the RMS of the instantaneous temperature contrast $\Theta \equiv \Delta T / \left< T \right>$, $\left< T \right>$ is the average photospheric temperature, and $\Delta T \equiv T - \left< T \right>$ is the difference from the mean of the photospheric temperature at any one location.
\citet{Samadi2013a} also provides the expressions
\begin{align}
	N_g &= \frac{2\pi R_s^2}{\Lambda^2} \label{eqn:N_g} \\
	\tau_g &= \kappa\rho\Lambda \label{eqn:tau_g},
\end{align}
where $R_s$ is the radius of the star, $\kappa$ is the Rosseland mean absorption coefficient, $\rho$ is the mean photospheric density determined as in Section~\ref{sec:mach}, and $\Lambda$ is a characteristic granular size\footnote{This characteristic size is used as a horizontal size in Equation \eqref{eqn:N_g} and as a vertical size in Equation \eqref{eqn:tau_g}. This is because both sizes are of the order of the pressure scale height. The horizontal case is discussed in Section \ref{sec:Lambda}; for further discussion of the vertical case, see \citet{Trampedach2011}, among others.} described in Section~\ref{sec:Lambda}.

With Equations \eqref{eqn:N_g} and \eqref{eqn:tau_g} along with the expression $R_s^2 = GM/g$, Equation~\eqref{eqn:sigma_tau} can be written as
\begin{align}
	\sigma &= 6 \sqrt{\frac{\kappa\rho\Lambda^3}{\pi R_s^2}}\;\Theta\RMS^2 \\
	&= \frac{6}{\sqrt{\pi}} \sqrt{\frac{\kappa\rho g}{GM}} \;\Lambda ^{3/2} \;\Theta\RMS^2. \label{eqn:sigma}
\end{align}
The subscript $\tau$ has been removed, as granulation is seen in a small region around a single optical depth of $\tau \sim 1$, and so the observed, total fluctuation amplitude $\sigma$ can be taken to be $\sigma_{\tau=1}$.
In past work, the expression for $\sigma$ was written in terms of the observable $\nu\max$, the peak frequency of p-mode oscillations assumed to scale as $\nu\max \propto g / \sqrt{T}$; however, we omit this step in the present work.
Additionally, \citet{Samadi2013b} compared their version of this expression to $\sigma$ values measured in numerical models and found the fit could be slightly improved by raising the expression to the power 1.10.
However, we omit this step to to remain closer to a model derived from first principles.
The present expression for $\sigma$ differs from past work most notably in that it is presented in absolute form, rather than as a scaling relation normalized to a solar $\sigma$ value.

\subsection{Granular Size}
\label{sec:Lambda}

\citet{Samadi2013a,Samadi2013b} define the characteristic granular size $\Lambda \equiv \beta H_p$ as proportional to the pressure scale height $H_p$, which is itself a function of \Teff\ and \logg.
The proportionality constant $\beta$ is a free parameter of the model.
By comparison of the modeled and observed power spectra of granular flicker, appropriate values of $\beta$ are shown to fall approximately in the range 3--15, depending on how the modeled spectra is constructed.
(This range is in agreement with the grid of simulations of \citet{Magic2013}, which finds this proportionality constant to be $\sim 5$.)

In this work, we instead use the $\Lambda(\Teff,\logg)$ scaling of \citet{Trampedach2013a},
\begin{equation}
	\log \frac{\Lambda}{[\text{Mm}]} \simeq 1.3210\;\log \Teff - 1.0970\;\logg + 0.0306,
	\label{eqn:Lambda}
\end{equation}
which those authors produced by fitting granular size measurements taken from a grid of numerical simulations.
They note that this functional form provides a better fit than a simple proportionality with $H_p$, as they find differing best-fit values of $\beta$ to be appropriate across the H-R diagram, with values ranging from 9 to 13 and generally increasing from dwarf stars to more evolved stars.
We note that a similar trend can be found in the \kep\ \fe\ measurements: if we take the final and completed model for \fe\ described through this section but use $\Lambda = \beta H_p$, and if we divide the \kep\ sample into 2D bins in \Teff--\logg\ space and determine the value of $\beta$ that minimizes the RMS error of model-predicted \fe\ values within each bin, then the resulting best-fit $\beta$ values are near 8--10 for main-sequence stars and rise to 12--20 for giants, echoing the trend in $\beta$ values observed in the simulations of \citet{Trampedach2013a} and supporting the use of the alternative functional form of Equation~\eqref{eqn:Lambda}.

Some solar observations have found two populations of granules \citep[e.g.][]{Abramenko2012}.
The granule sizes of \citet{Trampedach2013a} are measured by finding the peak of the 2D spatial power spectra of granular images, and so their analysis includes both granular populations.
These two populations are divided in size: the large granules have diameters in a Gaussian distribution of approximately $1.2\pm0.5$~Mm and are believed to be traditional convective cells, whereas the smaller granules (or granule-like visible features) follow a decreasing power-law distribution in diameter, with typical sizes under 0.5~Mm.
This population of smaller features may have its origin in turbulent processes at the solar surface rather than convection \citep{VanKooten2017}, meaning they may follow different distributions in timescale $\tau_c$ and temperature contrast $\Theta\RMS$ than those assumed in the present work's model.
\citet{Abramenko2012} quantify the importance of a given size of granule to the overall appearance of the photosphere with the area contribution function (ACF), defined as the ratio of the total area of granules of a given size to the total area available, and this should serve as a good proxy of the influence of granules (or granule-like features) of a given size on \fe.
While those authors report a bimodal ACF distribution, with a peak in equivalent diameter near the typical granular size of 1.2~Mm and a second peak near 0.5~Mm, we note that when the ACF is computed using their fitted size distribution, which does not deviate significantly from the observed distribution, it produces only a single peak near 1.2~Mm.
We are thus unable to make a clear determination on the degree, if any, to which this population of small, granule-like features contributes to \fe.
In this work we do not attempt to account for these smaller features.

We also note that the scaling of \citet{Trampedach2013a} produces a value for solar parameters $\left(\Teff=5770~\textrm{K}, \logg=4.438\right)$ of 1.35~Mm, which is in good agreement with observational values of the typical solar granular size (e.g. the range around 1.2 Mm of \citet{Abramenko2012}).

\subsection{Determining the Temperature Contrast}
\label{sec:theta}

An expression for $\Theta\RMS$ must be determined before this model can be used.
Mixing length theory (MLT) predicts that $\Theta\RMS$ is proportional to the square of the Mach number \Ma.
Motivated by this, \citet{Samadi2013b} fit a quadratic polynomial to the $\Theta\RMS$ and \Ma\ values measured in a grid of numerical simulations.
We note, however, that their fitted polynomial is a concave-down parabola (i.e. $\Theta\RMS \propto -\Ma^2$), as opposed to the concave-up parabola (i.e. $\Theta\RMS \propto \Ma^2$) expected by MLT.
Additionally, the simulation measurements themselves that were used for this fit appear to show a possible trend of flattening off at larger \Ma\ values, despite a general adherence to an approximate $\Theta\RMS \propto \Ma^2$ scaling for low \Ma.

\begin{figure}[tp]
	\centering
	\includegraphics[width=\linewidth]{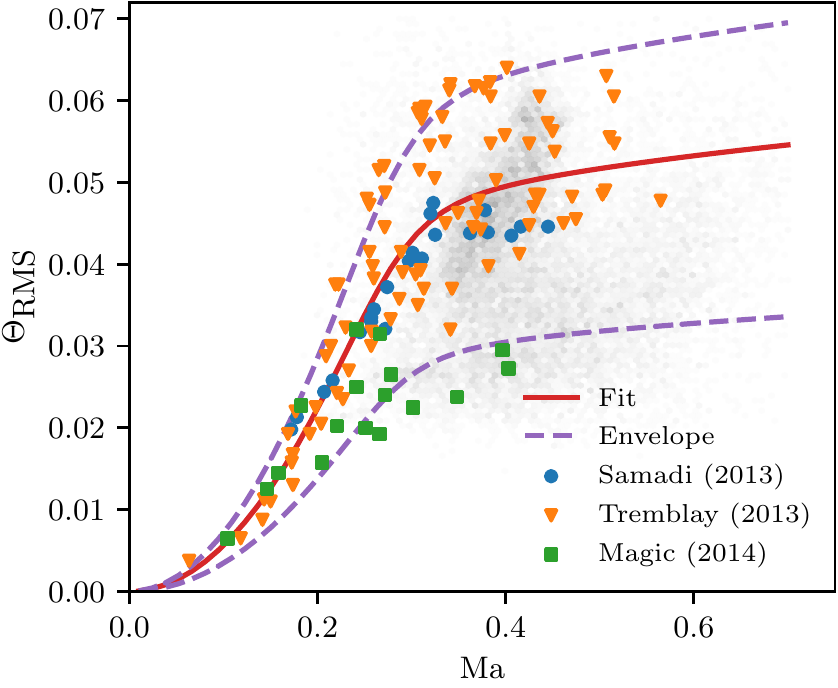}
	\caption{$\Theta\RMS$ as a function of Mach number \Ma. Dots mark measurements from the three sources of numerical simulations (see Section~\ref{sec:theta}). Lines mark our central fit to all points as well as our envelope fit to the upper and lower bounds of the point cloud. The faint gray background is a 2D histogram of the Kepler stars (of Section~\ref{sec:data}) provided for reference, plotting the value of $\Theta\RMS$ required for our model to reproduce the observed \fe, and \Ma\ as computed from Equation~\eqref{eqn:ma}.}
	\label{fig:ma-theta}
\end{figure}

For this work, we use an expanded data set, with $\Theta\RMS$ and \Ma\ measurements from additional simulations \citep{Tremblay2013,Magic2014}, shown in Figure~\ref{fig:ma-theta}.
The expanded set of simulations notably provides a wider range of $\Theta\RMS$ values that are consistent with any given \Ma\ value.
To account for this, we produce both a central fit to the data set as well as a fitted envelope encapsulating this spread.
The envelope allows the model to produce a range of possible $\sigma$ values for any given star, all consistent with the range of $\Theta\RMS$ values produced by the numerical simulations.
We use the functional form
\begin{equation}
	\Theta\RMS = \frac{1}{(A_1 \Ma ^ {-2c} + A_2 \Ma ^ d)^{1/c}},
	\label{eqn:theta-ma-fit}
\end{equation}
which is able to act as the $\Theta\RMS \propto \Ma^2$ predicted by MLT for small values of \Ma\ while more freely fitting the data points at higher \Ma, which may be beyond the applicability of ideal MLT.
We produce a central-fit curve $\Theta_\text{central}(\Ma)$, shown in Figure~\ref{fig:ma-theta}, by fitting the complete set of data points from all three numerical experiments, producing the coefficients $A_1=21.0$, $A_2=3.54\times10^6$, $c=5.29$, and $d=-0.842$.
To produce the upper bound of our envelope we identify by hand a set of points representing the largest values of $\Theta\RMS$ predicted for any given $\Ma$, and we find a constant scaling factor which, multiplying $\Theta_\text{central}$, best fits that subset of points.
This produces $\Theta_\text{upper}(\Ma)=1.27\;\Theta_\text{central}(\Ma)$.
Determination of the lower bound of the envelope is less clear, since there is a sharp transition in the lowest-reported $\Theta\RMS$ values near $\Ma=0.4$.
Following the same method as for the upper bound, while focusing on matching the low-$\Theta\RMS$ points for $\Ma>0.4$, produces $\Theta_\text{lower}(\Ma)=0.82\;\Theta_\text{central}(\Ma)$; focusing on the low-$\Theta\RMS$ points for $\Ma<0.4$ produces $\Theta_\text{lower}(\Ma)=0.62\;\Theta_\text{central}(\Ma)$.
We choose the latter option, which produces a larger envelope and is more inclusive of the range of $\Theta\RMS$ values seen in simulations.

While use of this envelope fit may seem ad-hoc, we believe it has some justification.
In Figure~\ref{fig:ma-theta}, the faint gray background represents an attempt to position our \kep\ star sample in the plot.
For these stars, the \Ma\ is that computed from Equation~\eqref{eqn:ma}, and the $\Theta\RMS$ value is that which would be required for our model to reproduce the star's observed \fe.
The spread seen in these empirical $\Theta\RMS$ values is very comparable to that seen in the simulations and which we capture with our envelope fit, meaning that this level of spread appears plausible (though its origin is unclear---see Section~\ref{sec:discussion}).

When viewing the \kep\ sample in Figure~\ref{fig:ma-theta}, it can be seen that the simulations do not span the full range of \Ma\ values that we produce for the \kep\ stars.
This mismatch is explainable by the fact that these simulation grids were not designed with our particular sample in mind.
The highest \Ma\ values for our sample correspond to the hottest stars, near 6500--7000~K, while the simulation grids stop at slightly cooler stars.
Despite this, the bulk of our \kep\ sample is well-covered by these simulations.

Using Mach numbers computed as described in the following section, our computed $\Theta\RMS$ values increase toward higher temperatures along the main sequence, in agreement with the numerical results of \citet{Salhab2018}.
As an additional test, in Appendix~\ref{appendix:hinode} we describe our own determination of a solar value for $\Theta\RMS$ from \textit{Hinode}/SOT observations.
We find a value of $\Theta\RMS=0.0457$, which exceeds some other observational values but does have some support in existing literature.
This value compares well with the value predicted by our central-fit model for solar parameters, which is $\Theta\RMS=0.0455$.

\subsection{Determining the Mach Number}
\label{sec:mach}

\begin{figure}[tp]
	\centering
	\includegraphics[width=\linewidth]{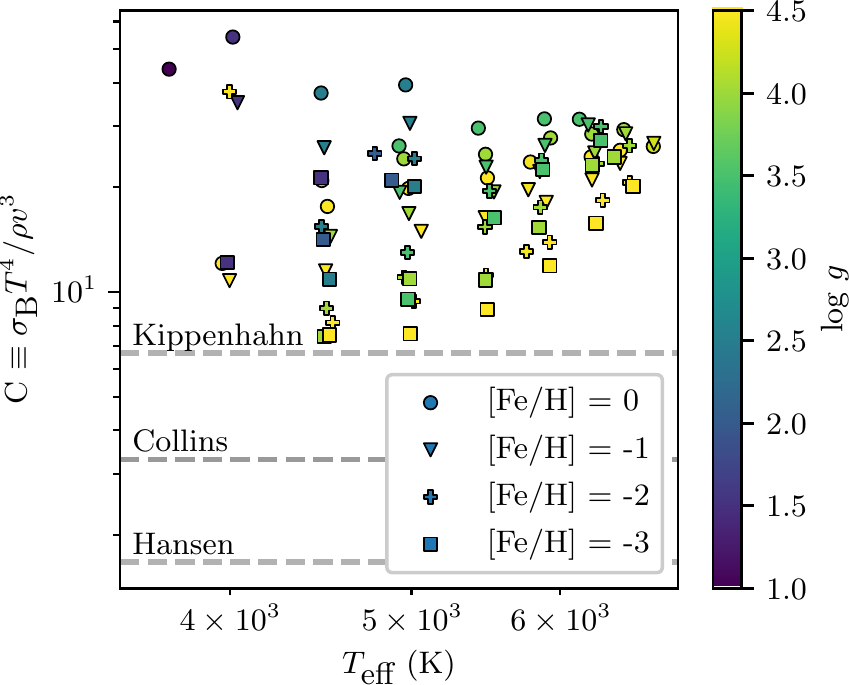}
	\caption{Computed $C$ values for the \citet{Tremblay2013} simulation grid. The grid stars have one of four discrete values of \feh. Also indicated are three $C$ values predicted by various formulations of mixing-length theory (see text).}
	\label{fig:C-cf-ml}
\end{figure}

\begin{figure*}[tp]
	\centering
	\includegraphics[width=\linewidth]{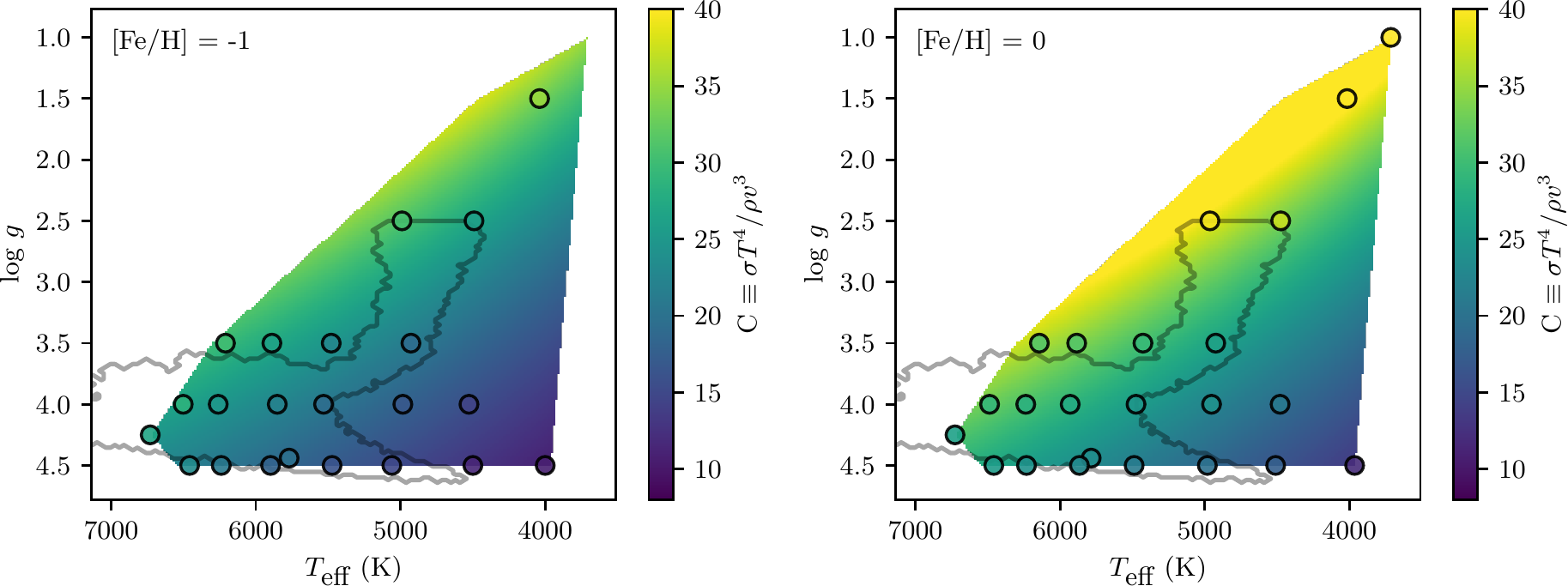}
	\caption{Our fitted function (Equation~\eqref{eqn:fitted-C}) for $C$ as a function of \Teff, \logg\, and \feh\ (see Section~\ref{sec:mach}). The colored, circular points represent the $C$ values calculated for simulations in the grid of \citet{Tremblay2013}. These simulations have one of only four discrete values of \feh\ (-3, -2, -1, and 0); we show here simulations and our fit for the two \feh\ values within the range of our dataset ($\feh=-1$ on left and $\feh=0$ on right), whereas all four are used when producing our fit. Interpolation and extrapolation through our fit produces $C$ values for our \kep\ stars, which have typical \feh\ values in the range $[-0.75, +0.5]$. The gray outline in these figures indicates the location of our \kep\ data set, provided as a quick reference.}
	\label{fig:C-fit}
\end{figure*}

A value for the Mach number \Ma\ (which we define in terms of the RMS vertical velocity $v$) must now be determined.
We start by assuming that some fraction of the total stellar flux is carried by convection, and we adopt a standard mixing-length theory expression for the convective flux:
\begin{equation}
	\sigma_{\rm B} \Teff^4 = C \rho v^3,
\end{equation}
where $\sigma_{\rm B}$ is the Stefan-Boltzmann constant, $\rho$ and $v$ are the plasma density and RMS vertical plasma velocity, and $C$ is a constant of proportionality.
Density $\rho$ is a function of \Teff, \logg, and metallicity $Z$, and the dependence on metallicity is an addition over the model as employed by \citet{Cranmer2014}.
We determine density \citep[following][]{Cranmer2011} by finding an interpolated value for the Rosseland mean opacity $\kappa_\textrm{R}$ from an AESOPUS data grid \citep[see][]{Marigo2009}, computing a photospheric value for the density scale height $H_\rho$ as in \citet{Cranmer2011}, and solving for $\rho$ after setting the photospheric optical depth $\tau = \kappa_\textrm{R}\rho H_\rho = 2/3$.

The Mach number \Ma\ is then
\begin{equation}
	\Ma \equiv v / c_s = \frac{\left(\sigma_{\rm B} \Teff^4 / C \rho\right)^{1/3}}{c_s},
	\label{eqn:ma}
\end{equation}
where $c_s$ is the stellar surface sound speed defined as $c_s^2 = 5 k_B T / 3 m_H \mu(T)$, where $k_B$ is the Boltzmann constant, $m_H$ is the mass of a hydrogen atom, and $\mu(T)$ is the mean atomic weight for which we use the fitted expression
\begin{equation}
	\mu \approx \frac{7}{4} + \frac{1}{2} \tanh \left( \frac{3500 - \Teff}{600} \right)
	\label{eqn:mu}
\end{equation}
of \citet{Cranmer2011}.

We require \Ma\ near the stellar surface ($\tau=2/3$) where the MLT expression may not be fully valid, and we account for this through the $C$ parameter.
To determine this proportionality constant, we turn to the simulation grid of \citet{Tremblay2013}, for which the surface values for the plasma density and Mach number are reported.
This allows a value of $C$ to be inferred for each simulated star in the grid, which we show in Figure~\ref{fig:C-cf-ml}.
(We note that the reported Mach numbers are the RMS velocity amplitude.
We rescale these values by $1/\sqrt{3}$ to produce vertical-component Mach numbers.
This scaling is supported by our own analysis of $\tau=1$ slices from a \muram\ solar-surface simulation \citep[see][]{Rempel2014}, in which the space- and time-averaged velocity components $v_x$, $v_y$, and $v_z$ are in near-exact equipartition.)

Of note is that all of these computed $C$ values exceed those predicted by multiple MLT formulations.
In Figure~\ref{fig:C-cf-ml}, we mark three predicted $C$ values.
The first, derived from the formulation of \citet{Kippenhahn2012}, is $C=4/(\alpha \nabla_\textrm{ad})$, using the standard MLT quantities the mixing-length parameter $\alpha$ and the dimensionless adiabatic temperature gradient $\nabla_\textrm{ad}$.
With $\alpha=1.5$ and $\nabla_\textrm{ad} = 2/5$, this gives $C=6.67$.
The second, of \citet{Collins1989}, is $C=2/(\alpha \nabla_\textrm{ad})$, yielding $C=3.3$ using the values above.
The third, due to \citet{Hansen2004}, is $C=1/(\alpha \nabla_\textrm{ad})$, giving $C=1.67$.
The fact that all these MLT predictions fall short of our computed $C$ values may be due to the breakdown of MLT assumptions at the stellar surface, where convective flows must come to a halt.

We fit these $C$ values with the function
\begin{equation}
	C = 6.086\times10^{-4} \; \Teff^{1.406} \; g^{-0.157} \; \left( Z / Z_\odot \right)^{0.0975},
	\label{eqn:fitted-C}
\end{equation}
where \Teff\ is expressed in K and $g$ is expressed in cm~s$^{-2}$.
Figure~\ref{fig:C-fit} shows how this function compares with the simulation grid, and we go on to use this function to calculate $C$ values for all stars when computing Mach numbers.

As a point of reference, for solar parameters ($T=5770$~K, $\logg=4.438$) this fit produces $C=23.648$.
This corresponds to a solar \Ma\ of 0.32 (or 2.6~km~s$^{-1}$).
Observational values of vertical velocities at the solar photosphere range from 1 to 3~km~s$^{-1}$ \citep[][and references therein]{Oba2017a}, with the range in values possibly due in part to observational limitations, and possibly due in part to a strong gradient in the vertical velocity seen in numerical simulations near the photosphere \citep{Fleck2020}, raising the possibility that different observations sample different portions of this stratification.
Nonetheless, the predicted solar \Ma, while high, is consistent with at least some observations.
Since the prediction that $\Theta\RMS$ scales with $\Ma$ is based on mixing-length theory, which does not hold at the surface of the convection zone, it may be appropriate that our model predictions for $\Theta\RMS$ are driven by relatively higher \Ma\ values that correspond to layers slightly deeper than the photosphere where MLT holds more strongly.

We note that this \Ma\ calculation includes the effect of metallicity at two points, in the calculation of $\rho$ and $C$.
Over the range of \feh\ values in our catalog (approximately $-1 < \feh < 0.5$) and for otherwise solar parameters, increasing metallicity causes a decrease in $\rho$ by a factor of approximately 3 over the full range of metallicity values.
The Mach number, however, has a minimum near $\feh=-0.75$, with an increase of approximately 20\% up to $\feh=0.5$ but a much slower increase toward lower metallicity.
The corresponding effect on \fe\ is very comparable to that in \Ma.

\vfill\null

\subsection[Limiting to <8 hr Timescales]{Limiting to $<8$~hr Timescales}
\label{sec:sigma-to-f8}

Once a value of $\sigma$ (the RMS flicker amplitude over all timescales) is calculated, it must be converted to the observational \fe\ (the RMS flicker amplitude over the $<8$ hr timescales dominated by granulation).
We follow \citet{Cranmer2014}, which assumed a Lorentzian function for the granular power spectrum and derived the relation
\begin{align}
	\label{eqn:fe-sigma}
	\frac{\fe}{\sigma} &= \CBP \sqrt{1 - \frac{2}{\pi} \tan^{-1}\left(4\;\tau\eff\;\nu_8\right)},
\end{align}
where $\nu_8 \equiv \left( 8\;\;\text{hr} \right)^{-1}$.
The factor \CBP\ has been inserted in this work to represent our bandpass correction term (see Section~\ref{sec:bandpass}).
We follow \citet{Samadi2013b} in defining $\tau\eff$, the characteristic timescale of granulation, as $\tau\eff = \Lambda / v$, where $\Lambda$ is the characteristic granular size of Section~\ref{sec:Lambda} and $v = \Ma c_s$ is the characteristic vertical plasma velocity.
In the present work, $\Lambda$ and $\tau\eff$ can be computed directly, and so we are not required to compute $\tau\eff$ as a scaling relation relative to a solar value, as was done in past work.
Nevertheless, the model-predicted $\tau\eff$ for solar parameters is 8.8~min, in line with observational values for granular lifetimes in the range 7--9~min \citep{Nesis2002}.

Excluding \CBP, the conversion factor tends to be very close to 1 for dwarf stars, and tends toward 0.5 for giants.

\subsection{Bandpass Correction}
\label{sec:bandpass}

We add to our version of this model a \kep\ bandpass correction factor, given that our observational \fe\ values are \kep-derived.
At the conceptual level, this accounts for the fact that the blackbody spectra of different portions of a stellar photosphere (i.e. granules versus lanes) have different amounts of overlap with the \kep\ bandpass.
For cooler K-type stars with spectra nearer to the long-wavelength end of the bandpass, the bandpass will pass a smaller fraction of the light from the cooler lanes than from the relatively warmer granules, dimming the lanes relative to the granule centers and exaggerating the effective temperature contrast when viewed in terms of measured intensity.
Conversely, for warmer F-type stars nearer to the short-wavelength end of the bandpass, the warmer granules will be dimmed relative to the lanes, reducing the effective temperature contrast.
(We find the effect to be much more pronounced for K stars than F stars.)
Analytically, this effect alters the $B = \sigma_{\rm B} T^4$ scaling inherent in \citet{Samadi2013a}'s derivation leading to our Equation~\eqref{eqn:sigma_tau}, producing a much stronger dependence on temperature when a star's peak wavelength of emission is near either end of the bandpass.
In Appendix~\ref{appendix:bandpass} we derive the correction factor \CBP\, which takes values of $\sim1$ (for F-type stars) to $\sim2$ (for K-type stars) and is nearly monotonic in effective temperature (see Figure~\ref{fig:sigma_correction_factor}). 

\section{Results}
\label{sec:results}

In this section we compare the observed and model-predicted \fe\ values, describe how the changes we made to the model affect that comparison, and discuss possible ways to interpret the final ratio of these two values.

\subsection{Initial Comparisons}
\label{sec:initial-comp}

First, as a simple benchmark, we compute the quantity $\Delta$, defined as
\begin{equation}
	\log_{10} \Delta \, \equiv \, \rm{RMS} \left[ \log_{10} \left( \frac{\feobs}{\femod} \right) \right],
\end{equation}
which measures the typical multiplicative factor separating each star's observed and model-predicted \fe\ values.
(For example, a value of $2$ would indicate the typical star is either over- or under-predicted by a factor of $2$, with mispredictions in either direction contributing equally to the overall score.)
Calculated across all stars in our catalog, we find a value of $2.50$ when using the predictions of the model of \citet{Cranmer2014} (without their empirical correction factor), and a decreased value of $2.02$ when using the model presented in the present work.
Perfect model--observation agreement would produce a value of $1$, so this decrease represents a 32\% reduction in the typical, relative prediction error.
As an additional metric, if an ad-hoc correction factor were to be introduced that multiplies all model predictions, $\Delta$ is minimized when that factor is $0.66$, with $\Delta=1.75$.
(To be clear, we do not propose such a correction factor in this work.)

\begin{figure}[tp]
	\centering
	\includegraphics[width=0.95\linewidth]{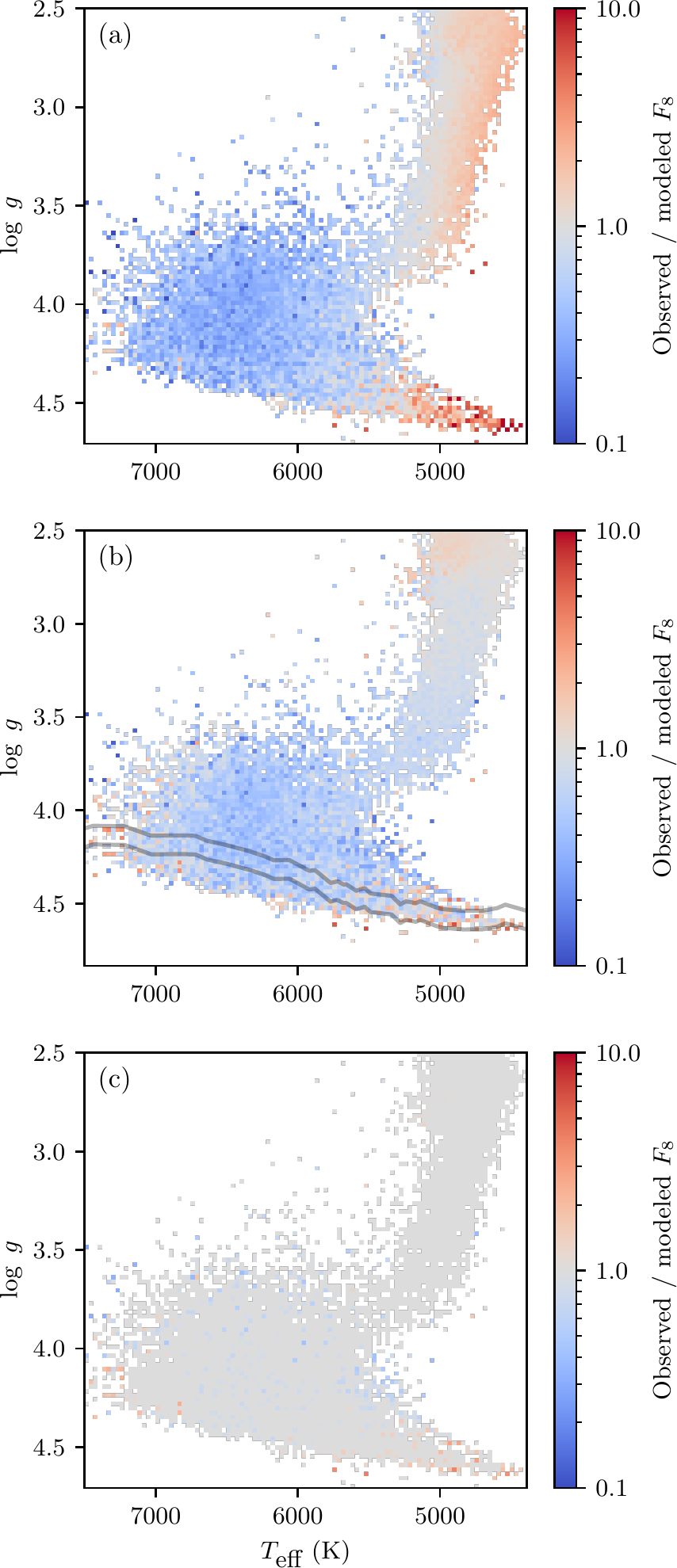}
	\caption{Two-dimensional histograms showing the median ratio of observed to modeled \fe, (a) using the model as in \citet{Cranmer2014} (without their empirical correction factor), (b) using our model, and (c) using our model with the full envelope of possible $\Theta\RMS$ values for each predicted \Ma. That is, the ratio is shown as 1 if any of the possible $\Theta\RMS$ values reproduce the median observed \fe\ in a histogram bin, and otherwise the ratio is between the observed \fe\ and the nearest of the possible model predictions. The dark lines in (b) mark the main-sequence slice we use in Figure~\ref{fig:ms-slice}.}
	\label{fig:before-and-after}
\end{figure}

\begin{figure*}[tp]
	\centering
	\includegraphics[width=\linewidth]{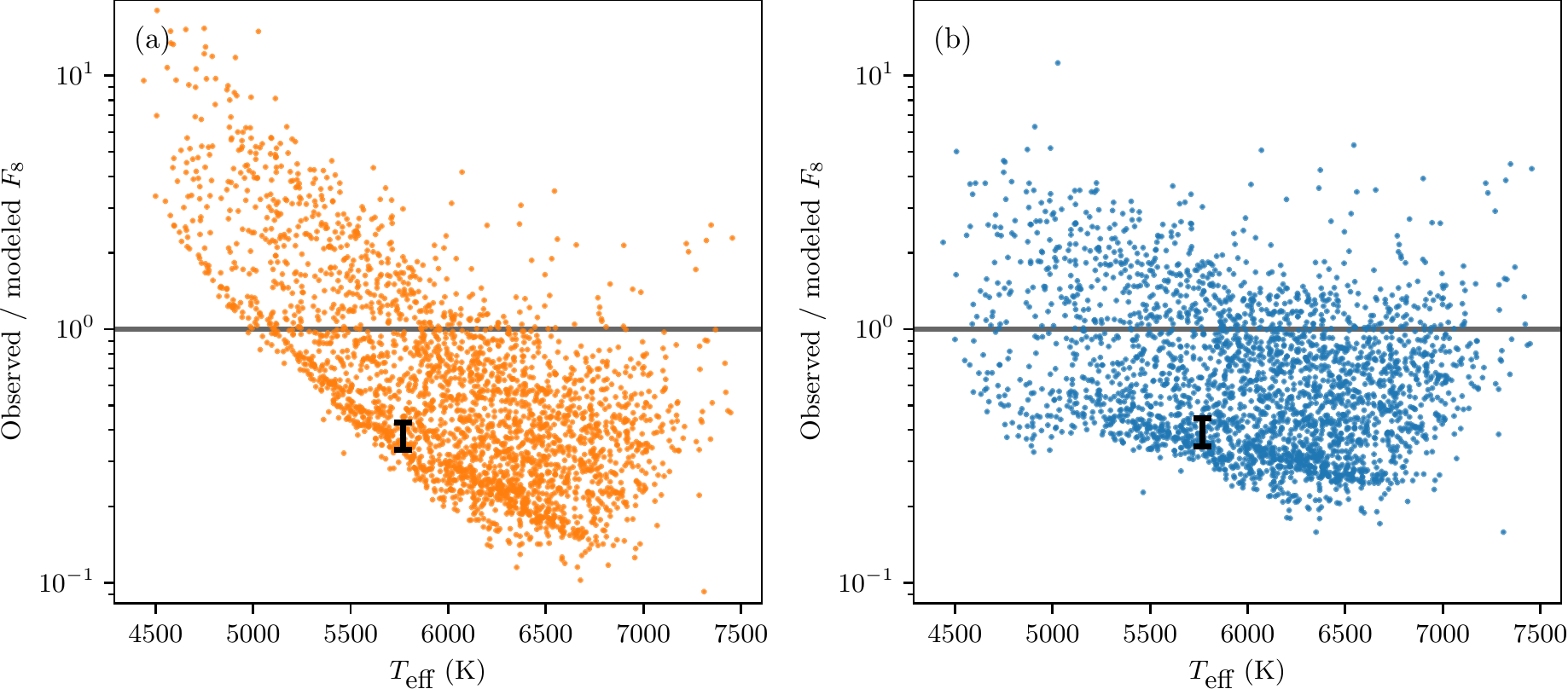}
	\caption{The ratio of observed to modeled \fe\ values along the main-sequence slice shown in Figure~\ref{fig:before-and-after}b. In (a) we use the model as in \citet{Cranmer2014} (without their empirical correction factor), which shows a strong trend with temperature, and in (b) we use our model (without the envelope fit for $\Theta\RMS$), showing that the temperature trend has been significantly mitigated. The black bars mark the Sun and its range of observed \fe\ values (discussed further in Section~\ref{sec:solar_value}).}
	\label{fig:ms-slice}
\end{figure*}

In Figure~\ref{fig:before-and-after} we show the ratio of observed to modeled \fe\ across stellar types, both before and after the modifications we have made to the model in the present work.
We also show the expanded agreement afforded by the ``envelope'' in the $\Theta\RMS(\Ma)$ fit (see Section~\ref{sec:theta}).
Considering first the central-fit model (comparing Figures \ref{fig:before-and-after}a and \ref{fig:before-and-after}b), the model agreement has improved almost universally, with significant improvement among the K-type dwarfs and the cool edge of the giant branch.
However, along the main sequence, the appearance of a small band of maximal model agreement near 5500~K has been reduced.
(This feature's appearance in the model of prior work may have been due to the handful of areas in which the model was calibrated relative to solar parameters, whereas our current model removes these close ties to the Sun.)

We turn now to the envelope fit in Figure~\ref{fig:before-and-after}(c), which accounts for the fact that simulations predict a range of possible $\Theta\RMS$ values for a given \Ma\ (producing an envelope around the $\Theta\RMS(\Ma)$ fit line; see Figure~\ref{fig:ma-theta}), leading to a range of possible \fe\ values.
The median observed \fe\ values for nearly all bins fall within the \fe\ range predicted by this envelope model, and so these bins appear in the plot with a ratio of 1.
On a star-by-star basis, 78\% of stars have an observed \fe\ falling within the appropriate envelope range.
Given the spread in predicted $\Theta\RMS$ values from numerical simulations, this envelope fit may be the more reasonable representation of the level of certainty available in modeling \fe.

Past work \citep{Cranmer2014} noted that F-type stars display less flicker than predicted in a manner that appeared temperature-dependent, and the prior model applied to our current, expanded sample of measured \fe\ values shows the opposite effect in the K-type dwarfs, yielding a strong, temperature-dependent trend along the main sequence.
We illustrate this more clearly in Figure~\ref{fig:ms-slice}, in which we plot $\feobs / \femod$ for the main-sequence slice indicated in Figure~\ref{fig:before-and-after}b.
(This slice is those stars within $0.05$~dex in \logg\ of the dwarf sequence values\footnote{Some of these values were originally presented in Table 5 of \citet{Pecaut2013}; we obtain updated values from version 2019.3.22 of \url{http://www.pas.rochester.edu/~emamajek/EEM_dwarf_UBVIJHK_colors_Teff.txt}} of \citet{Pecaut2013}.) 
Our updated model significantly mitigates that apparent trend, producing a more uniform model discrepancy along the main sequence.
Indeed, the remaining residual for our updated model trends more strongly (if at all) with \logg\ instead of \Teff.
\fe\ itself has a strong dependence on \logg\ (varying by an order of magnitude across $2.5 < \logg < 4.5$, see Figure~\ref{fig:flicker_data}) which is generally accounted for by the model.
It may be encouraging to see that the remaining discrepancy relates more closely to the primary variable \logg, and that the \Teff\ dependence present in past work is now resolved.

\begin{figure*}[tp]
	\centering
	\includegraphics[width=\linewidth]{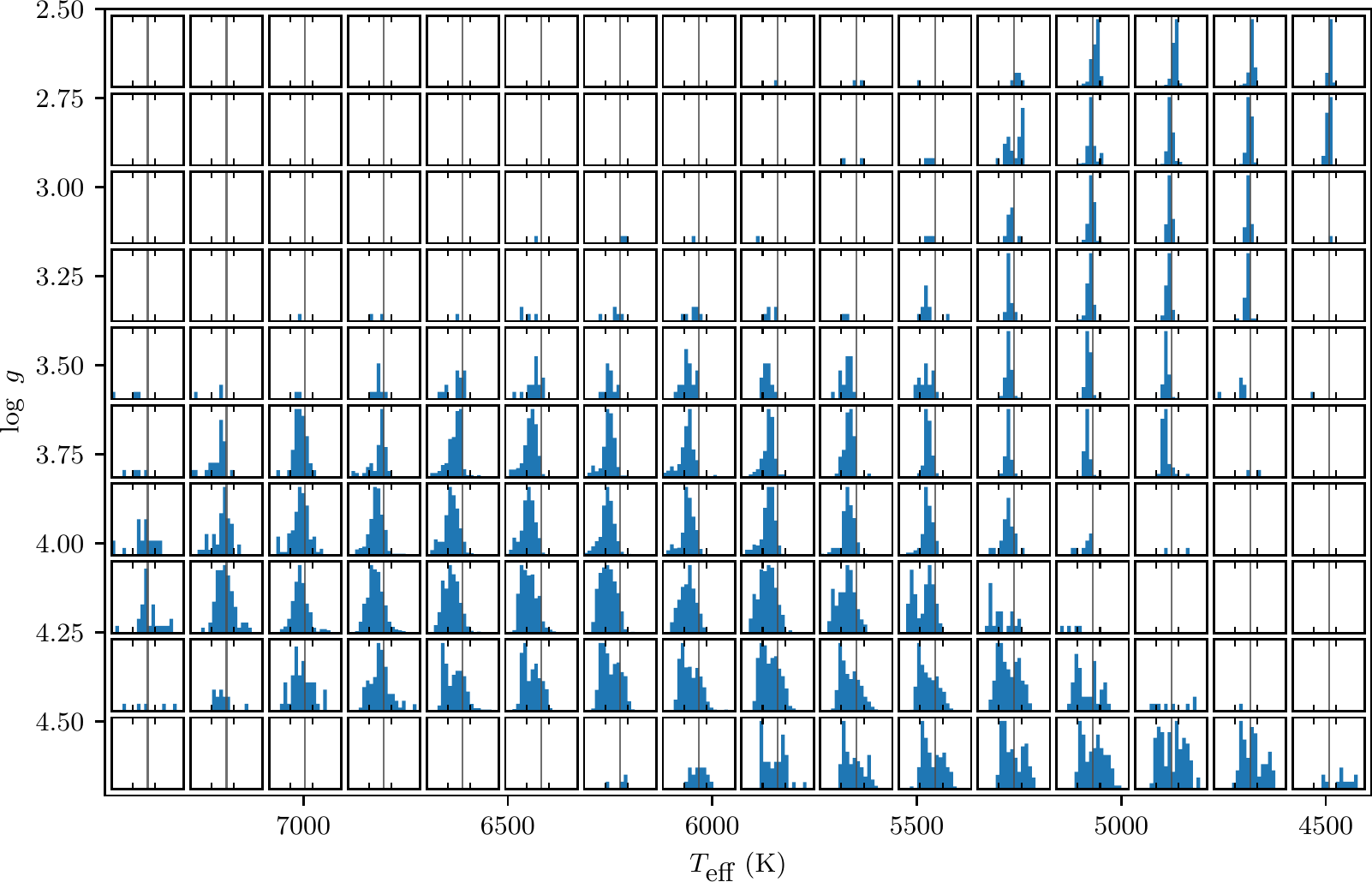}
	\caption{The distribution of $\feobs/\femod$ values. We have taken the $\Teff-\logg$ domain of Figure~\ref{fig:before-and-after} and divided it into a $10\times16$ grid. Within each grid cell, we plot a histogram of $\feobs/\femod$ for the stars within that cell. Each histogram's horizontal axis is logarithmically-scaled and runs from 0.1 to 10. A vertical, gray line marks 1, and small ticks mark the range afforded by our envelope fit (see text). Each cell's vertical axis is scaled independently to match the histogram's maximum value, but each vertical axis is required to span at least 0 to 10 so that cells with few stars can be identified visually.}
	\label{fig:mini-dists}
\end{figure*}

Figure~\ref{fig:before-and-after} compares only the median observed \fe\ value in each bin to model predictions.
In Figure~\ref{fig:mini-dists}, we display the distribution of $\feobs/\femod$ values (using our central-fit model) across the sample of stars.
When discussing these ratios, the ideal value is $1$ for every star, and since our envelope fit produces a constant multiplier for $\Theta$, with $\fe \propto \Theta^2$, the envelope produces a fixed range of explainable ratio values (from 0.38 to 1.61), indicated by the small tick marks in each sub-plot.
It can be seen that there exists a spread of observed \fe\ values relative to model-predicted values, especially nearer to the main sequence.
The typical spread of values is approximately symmetric (in logarithmic space) around some central value, and that central value is often near 1 and in almost all cases is within the envelope range.
However, in most of the cells in our plot, the central value is slightly below 1, in correspondence with the typical ratio values seen in Figure \ref{fig:before-and-after}b.
Visualizing these distributions provides an important reminder that the model predictions are often the  most accurate for the average star within any one cell.
While some of this spread is certainly due to measurement uncertainty, some of it seems to be real variation from star to star (as we illustrate for the Sun in Section~\ref{sec:solar_value}).
Of note is that the distribution of $\feobs/\femod$ values is more narrow for giant stars.
Recalling that \fe\ is larger by an order of magnitude for giant stars as compared to dwarfs, this might be explained by some fixed amount of variation in \feobs\ (whether arising from the star itself or from observational noise) causing a much smaller relative variation in \feobs\ for giants.

\subsection{On the Solar Value}
\label{sec:solar_value}

\begin{figure}[tp]
	\centering
	\includegraphics[width=\linewidth]{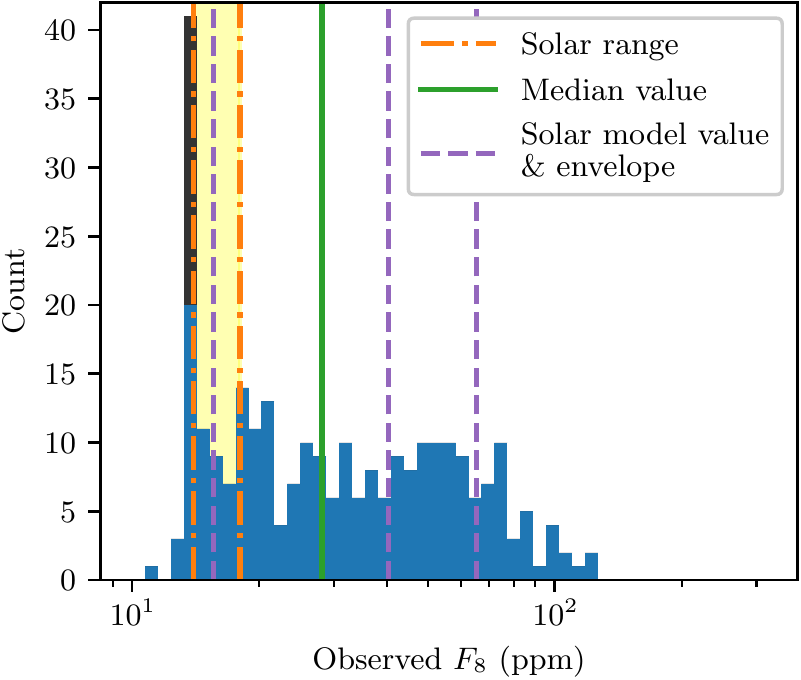}
	\caption{Histogram of 273 \kep\ stars in the ``solar neighborhood'' (5620~K $<\Teff<$ 5920~K, 4.363 $<\logg<$4.513). Vertical lines mark the median \fe\ of the \kep\ stars, the central-fit model-predicted \fe\ and the bounds of the envelope prediction for solar parameters, and the range of the solar \fe\ over its 11-year magnetic cycle. The excess of stars in the largest bin is due in part to a clamping applied by \citet{Bastien2016} during shot-noise removal when computing the measured \fe\ values, with the affected stars indicated by the darker portion of the histogram bar---many of those stars likely should be spread over even lower \fe\ values. (Such a clamping is present only in a narrow band of $T-\logg$ bins along the main sequence.)}
	\label{fig:sun_hist}
\end{figure}

Of interest is a careful look at the solar value of \fe, since portions of our model can be independently compared with solar observables.

First, we look at the solar value of \fe\ itself.
The Sun's \fe\ can be computed from SOHO/VIRGO light curves \citep{Basri2013}, and it ranges from 14 to 18~ppm, a range that includes the variation due to the 11~yr solar cycle \citep{Bastien2013}.
(These values are very similar to the \fe\ values computed by \citealp{Sulis2020}.)
In Figure~\ref{fig:sun_hist} we show this range in a histogram of \fe\ values for \kep\ stars within 150~K in \Teff\ and 0.075~dex in \logg\ of the Sun.
Of the 273 stars in the histogram, 71\% have a measured \fe\ greater than the upper end of the solar range (i.e. 18~ppm), and 85\% exceed the lower end of the solar range (i.e. 14~ppm).
The solar \fe\ also straddles the edge of the range allowed by our envelope fit.

A similar relationship between the Sun and the \kep\ sample was shown by \citet{Gilliland2011} using the Combined Differential Photometric Precision (CDPP) metric, which measures the total noise budget in \kep\ observations, including the ``intrinsic'' noise of the star itself over 6.5~hr timescales (i.e., flicker in the star's light curve).
Those authors find that only 23\% of solar-type \kep\ stars have a CDPP as low as the Sun, giving the Sun a position in the CDPP distribution comparable to its position in the \fe\ distribution for Sun-like stars.
In an alternative measure of photometric variability, the total range of variability (which is tied to magnetic activity but is not strongly influenced by granulation), the Sun is more typical compared to Sun-like stars \citep{Basri2013} though less active than those with detectable, Sun-like periodicity \citep{Reinhold2020}.

This difference between the Sun and most of the sample of Sun-like stars suggests that the Sun's granulation pattern, and perhaps the convection driving it, may vary to some degree from that of a typical Sun-like star.
Within this sub-sample, neither the observed \fe\ values nor the ratio of observed and model-predicted \fe\ values correlates strongly with \Teff, \logg, \feh, rotation rate (for the 72 stars with measured periods) or magnetic activity indices (for the 53 stars with measured indices).
(Rotation rates and magnetic indices are discussed further in Section~\ref{sec:addl-quantities}.)
Additionally, model values such as the characteristic size ($\Lambda$), temperature contrast ($\Theta\RMS$), and timescale ($\tau\eff$) of granulation for solar parameters compare well to observed solar values (as shown in Sections \ref{sec:Lambda}, \ref{sec:theta}, and \ref{sec:sigma-to-f8}).
Despite this, the model-predicted \fe\ for solar parameters agrees well with the population median of Sun-like stars but not the Sun itself.
We discuss this disagreement further in Section~\ref{sec:discussion}.

This variation also enhances the value of the model presented in this paper, which is provided in more absolute terms, whereas past iterations included some scaling relations relative to solar values.
It should be noted, though, that past versions normalized their $\sigma$ (and therefore \fe) predictions not to the observed solar $\sigma$, but to the $\sigma$ of the solar-like simulation in the grid of \citet{Samadi2013b}.
The latter value, converted to \fe\, is in fact very close to the average \fe\ observed in the \kep\ sample for Sun-like stars.

\newpage

\subsection{Additional Observables Considered}
\label{sec:addl-quantities}

Here we describe a few observables which we considered as factors which might explain the remaining discrepancy between model predictions and observations (i.e. the quantity $\feobs/\femod$), but which we did not find to be useful.

\subsubsection{Rotation Rate}

\begin{figure*}[tp]
	\centering
	\includegraphics[width=\linewidth]{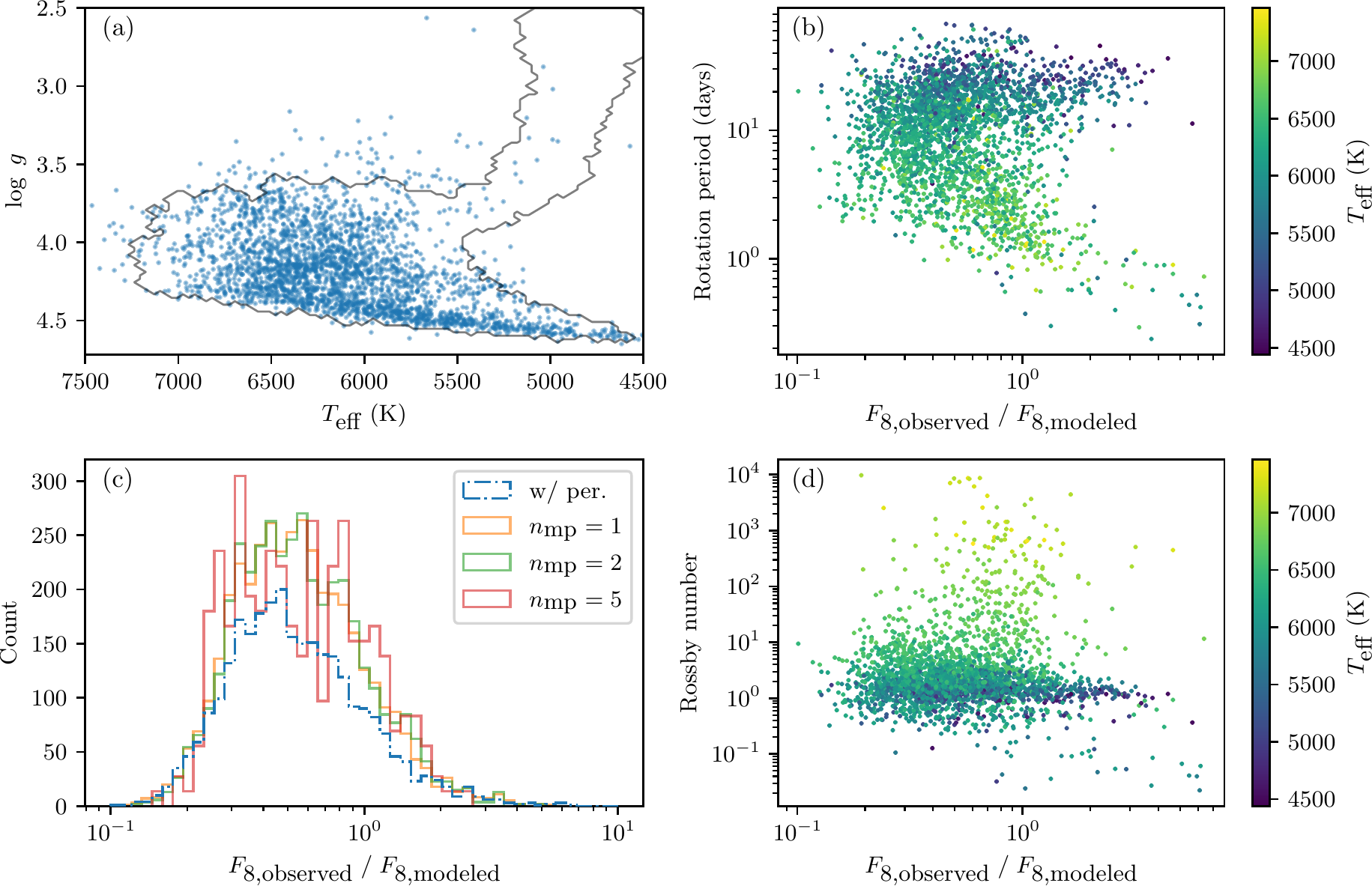}
	\caption{(a) Coverage of the \citet{McQuillan2014} sample. Blue dots mark stars included in both the rotation rate sample and our joint flicker-metallicity sample. The gray line marks the boundary of the bulk of our complete flicker-metallicity sample. (b,d) Distribution of periods or Rossby number with \fe\ model discrepancy. No clear correlation is seen, though we note the fastest-rotating stars tend toward increased \fe\ as a portion of the rotational signal leaks into the \fe\ window. (c) Histograms of the distribution of model discrepancies for stars with detectable periods (a marginalization of (b)) and similar stars, under varying criteria, without detectable periods (see text). The $n_\text{mp}=2,5$ histograms are scaled vertically to align with the $n_\text{mp}=1$ curve.}
	\label{fig:rot_rate}
\end{figure*}

\citet{McQuillan2014} provide rotation rate measurements of \kep\ main-sequence stars derived via the auto-correlation method.
These data cover 3,954 of the 27,628 stars in the flicker sample and 2,820 of the 16,992 stars in our joint flicker-metallicity sample (shown in Figure~\ref{fig:rot_rate}a).
While this sample does not include any giant stars, it covers most of the $\logg > 3.5$ portion of our \fe\ sample, which contains the bulk of the remaining model disagreement (as seen in Figure~\ref{fig:before-and-after}b).

$\feobs/\femod$ shows no clear trend with rotation period or Rossby number (calculated using Equation (36) of \citet{Cranmer2011}), suggesting that the remaining discrepancy in the model predictions is not due to rotation rate effects (Figure~\ref{fig:rot_rate}b,d).
The correlation coefficient of $\log\left(\feobs/\femod\right)$ with the logarithm of the rotation period is $r=-0.11$, and with $\log\left(\text{Rossby number}\right)$ is 0.025.
Notably, though, the fastest-rotating stars (periods $\lesssim 1$~day) do show an enhanced \fe, as the rotational power spectrum begins to cross into the 8~hr window of \fe.

Additionally, the model discrepancy does not vary strongly between stars with and without detected rotation periods.
(Such a trend could occur if stellar properties that affect the detectability of a rotation period correlate with flicker.)
To determine this, we divide the ($\Teff$, \logg) parameter space plotted in Figure~\ref{fig:rot_rate}a into a $100\times100$ grid (the same grid used, e.g., for binning in Figures~\ref{fig:flicker_data} and~\ref{fig:before-and-after}) and identify those stars without measured periods which fall within a grid cell containing at least $n_\text{mp}$ stars with a measured period.
This is to ensure similar sample populations for stars with and without measured periods, since stars with measured periods span a smaller range on the H-R diagram than our full \fe\ sample.
We repeat this for $n_\text{mp}=1,2,5$, which strikes different balances between covering the full extent of the stars with measured periods and covering only the core of this population, and we see no difference in the results.
Shown in Figure~\ref{fig:rot_rate}c, the population-matched sample of stars without measured periods shows a slightly flatter distribution, but the distribution of ratios is otherwise similar in location and extent for stars with and without measured periods.

Within the ``solar neighborhood'' of Section~\ref{sec:solar_value}, these results also hold.
Little correlation is seen between either rotation period or Rossby number and $\feobs/\femod$ (with Pearson $r=-0.080$ and $-0.068$, respectively), and no clear difference is seen between the distributions of $\feobs/\femod$ for stars with and without measured periods.

\subsubsection{Magnetic Activity}

\begin{figure*}[tp]
	\centering
	\includegraphics[width=\linewidth]{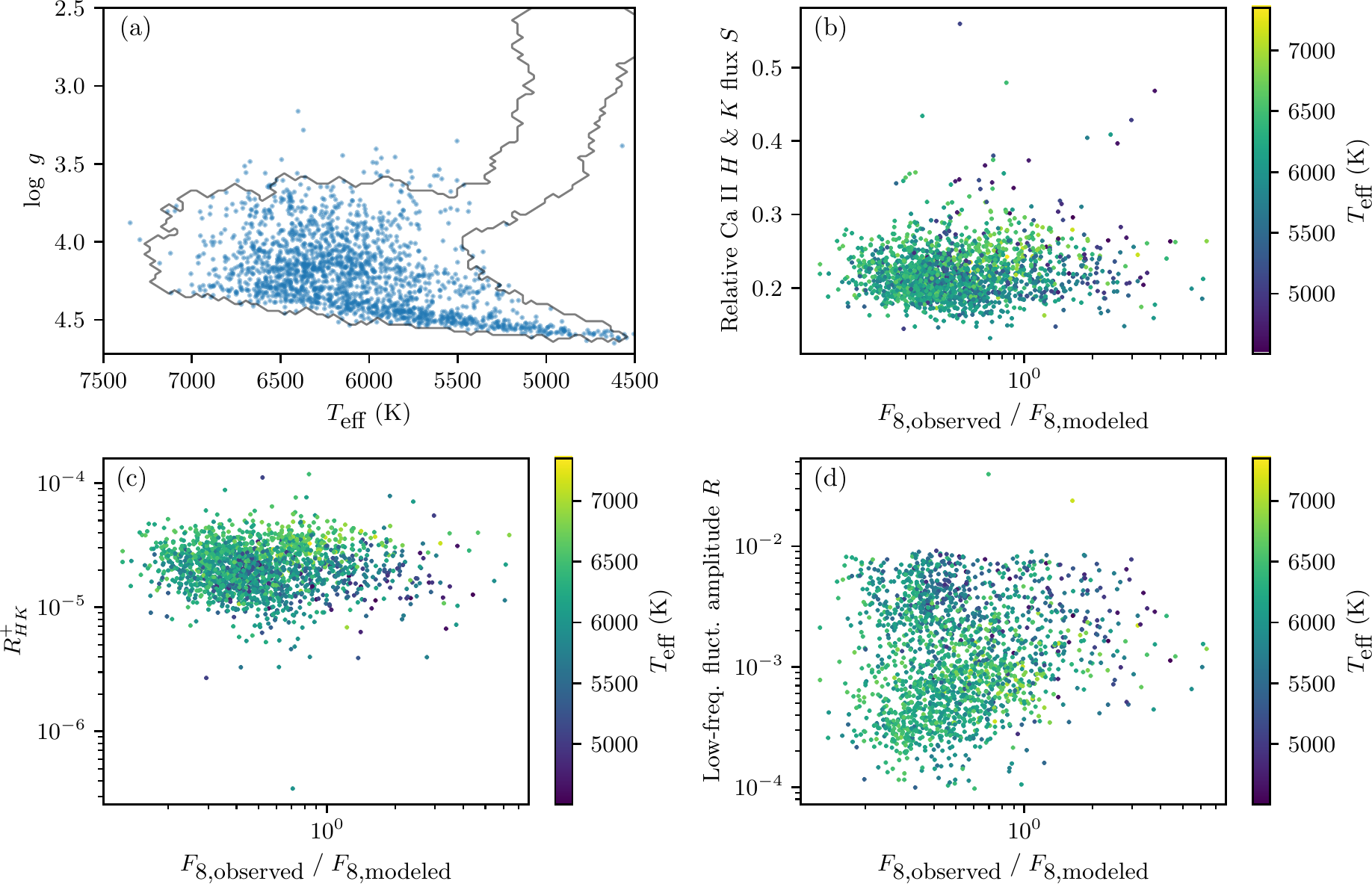}
	\caption{(a) Coverage of the \citet{Zhang2020} sample. Blue dots mark stars included in both the magnetic indices sample and our joint flicker-metallicity sample. The gray line marks the boundary of the bulk of our complete flicker-metallicity sample. There is no strong correlation between the ratio of observed to modeled \fe\ values and (b) relative Ca II $H$ and $K$ flux $S$, (c) $R^+_{HK}$, or (d) the relative low-frequency fluctuation amplitude $R$. The correlation coefficients, respectively, are $r=0.12, -0.011$, and $0.115$ (calculated using the logarithm of the latter two quantities).}
	\label{fig:mag_indices}
\end{figure*}

\citet{Zhang2020} provide measurements of the magnetic activity indices $S$, the ratio of emission in the Ca~II $H$ and $K$ lines to the continuum; $R^+_{HK}$, a proxy derived from $S$ by eliminating the contributions to the Ca~II $H$ and $K$ flux of the photosphere and the basal chromospheric flux; and $R_\text{eff}$, the range of low-frequency fluctuation in the \kep\ light curves of stars (i.e. the range of fluctuation due to magnetic activity, not the fluctuation due to granular activity considered in this paper).
$S$ and $R^+_{HK}$ are measured from LAMOST spectra.

The sample covers 2,021 stars with measured \fe\ and 1,895 stars in our joint flicker-metallicity sample.
As with the rotation rate sample, this sample covers most of the $\logg > 3.5$ portion of our \fe\ sample.
As shown in Figure~\ref{fig:mag_indices}, none of the three magnetic activity indices show any strong correlation with the ratio $\feobs/\femod$, indicating that the remaining discrepancy in the model predictions is not due to the types of magnetic activity measured by these indices (i.e., stellar magnetic flux in the chromosphere, and starspots and faculae on the photosphere).
This is in line with the findings of \citet{Meunier2017} that, while stellar magnetic activity does correlate with reduced convective signals, that effect is uniform across G and K dwarfs (whereas our model--observation discrepancy varies from early-G to K dwarfs).

In the solar neighborhood of Section~\ref{sec:solar_value}, only small, insignificant correlations are seen with these quantities.

\subsubsection{Binary Stars}
\label{sec:binaries}

\begin{figure}[tp]
	\centering
	\includegraphics[width=\linewidth]{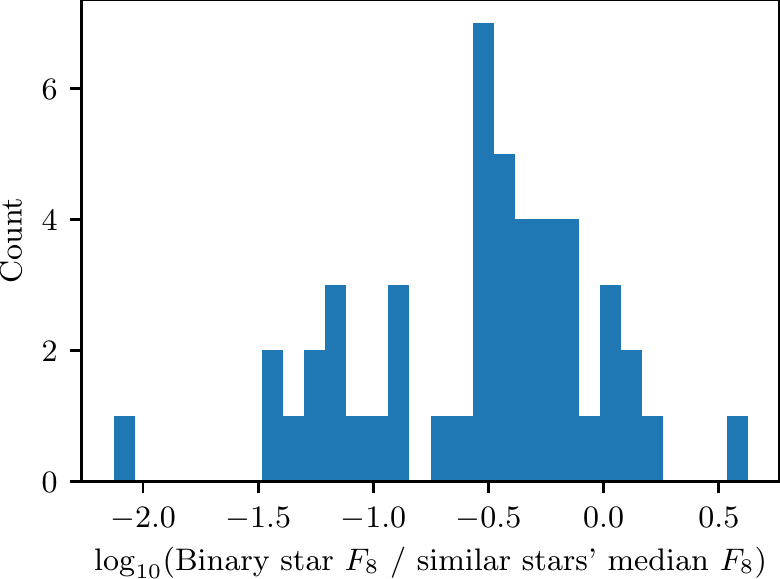}
	\caption{Histogram of the ratio of a binary star's observed \fe\ to the median of stars with similar \Teff\ and \logg, for each of 48 known binary stars.}
	\label{fig:binaries}
\end{figure}

\citet{Kirk2016} provide a catalog of 2,878 identified eclipsing and ellipsoidal binary stars in the \kep\ field (though we use the third revision of their online catalog\footnote{\url{http://keplerebs.villanova.edu/}}, which is expanded to 2,922 binaries).
This catalog includes 73 known binaries included in the flicker sample, and 48 known binaries in the joint flicker-metallicity sample.
This can by no means be considered a complete catalog of the binary stars in our sample, but it does allow the properties of known binaries to be compared to those of a mixed binary/non-binary population.
For each of the 48 known binaries in our sample, we identified all stars with a \Teff\ within 100~K and a \logg\ within 0.05~dex of the binary star (which will include a mixture of unknown binary stars and non-binary stars).
This selection produces of order 100 ``neighbor'' stars for each known binary.
We computed the ratio of each binary star's observed \fe\ to the median \fe\ of its neighbors, and these ratios are shown in Figure~\ref{fig:binaries}.
The geometric mean of this ratio across all 48 known binaries is 0.30, with nearly all values falling below one.
While the sample size is low, this seems to suggest that binary stars (or at least those binary stars that are easiest to detect) display lower \fe\ than non-binaries (since the ``similar stars'' comparison is a mixture of unknown binaries and non-binaries).

This reduction in mean \fe\ aligns with the fact that, in the case of an unresolved binary system, the presence of a binary companion star will cause the observed light curve to be the sum of the light curves of the two individual stars. Both stars' granules will therefore contribute to the variability of the combined light curve (along with a more difficult-to-quantify change to the effective values of the timescale and temperature contrast of granulation when viewing these commingled granular patterns), and $\fe \propto 1 / \sqrt{N_g}$.
This decrease in \fe\ may be able to explain some of the over-prediction of \fe\ by our model for most non-giant stars.
While our sample of binaries is small, the estimated multiplicity fraction for FGK dwarfs is of order 50\% \citep{Raghavan2010} (though we note the \kep\ sample of stars may be biased against binary stars; see \citet{Wolniewicz2021}), meaning binarity could be a factor for many stars.
In addition, we showed in Section~\ref{sec:initial-comp} that multiplying all model-predicted \fe\ values by 0.66 minimizes one metric of overall observation-model disagreement.
This factor is intriguingly close to the factor of $1/\sqrt{2}=0.71$ that would arise under the very simple assumption that the presence of a binary companion simply doubles the number of visible granules, with no other changes.

However, we emphasize that the very small number of identified binaries in our flicker-metallicity sample makes it difficult to draw any strong conclusions about the impact of binarity on \fe.
We feel that this may be an interesting topic for additional study with a larger sample of known binaries, but we do not attempt to incorporate binarity into the present work.


\section{Discussion}
\label{sec:discussion}

We have improved the ability of the theoretical model to predict \fe, but some discrepancies with the \kep\ data remain.
An important fact to note is that our model--observation comparisons are based on the median observed \fe\ within a \Teff--\logg\ bin, and that the population of stars within each bin shows a non-trivial degree of spread (as illustrated in Figure~\ref{fig:mini-dists}, and for Sun-like stars in Figure~\ref{fig:sun_hist}).
To some degree this is certainly due to measurement uncertainty in \fe.
However, the observed solar \fe\ is relatively well-constrained and shows relatively strong deviation from the typical \fe\ for Sun-like stars, suggesting a level of intrinsic variation in \fe\ between stars.
Thus, eliminating the model error seen in Figure~\ref{fig:before-and-after} is only the first step in fully modeling stellar flicker.

A number of possible avenues might resolve the remaining model errors.

The \fe\ model includes a handful of quantities (such as the characteristic size $\Lambda$, temperature contrast $\Theta\RMS$, plasma flow velocity \Ma, and timescale of granulation $\tau\eff$) which certainly vary across the H-R diagram, and this variation is characterized through numerical simulations.
Further numerical work to develop increasingly accurate scaling relations for these quantities will directly improve the \fe\ model.
In particular, an improved scaling for the granular temperature contrast $\Theta\RMS$, as well as for the Mach number on which $\Theta\RMS$ is expected to depend, may help explain the spread in $\Theta\RMS$ values seen in Section~\ref{sec:theta} and may eliminate the need for our envelope fit, or at least provide a more theoretically-grounded replacement for it.
Conversely, to the degree that any one of these quantities' scaling behavior is less well understood than the others, our \fe\ model may provide a route for observationally constraining that quantity for future simulations.

Beyond better constraints for factors already incorporated in the model, additional factors might need to be included.
One such factor is the presence of binary companions, discussed in Section~\ref{sec:binaries}.

Another possible factor is the effect of star spots.
While the 8~hr filtering applied to calculate \fe\ removes the direct influence of starspots on flicker (as their evolution and their rotation across the disk of the star both occur on longer timescales for all but the fastest-rotating stars), it is possible that spots still produce a small effect in \fe.
One mechanism might be the suppression of granulation within the boundaries of a spot, which reduces the number of granules contributing to \fe\ and thereby increases \fe.
The solar \fe\ has been shown to be independent of the solar cycle \citep{Bastien2013,Sulis2020}, suggesting that activity levels up to solar levels do not affect \fe.
We have not attempted to model this possibility for greater activity levels in the present work.

Another possibility lies in the influence of smaller-scale magnetic fields, including weaker, background fields, on convection.
\citet{Cranmer2014} presented an ad-hoc, empirical factor premised on the idea that F-type stars, with shallower convective layers, may be more susceptible to magnetic fields that constrain plasma and reduce convective velocities (an idea supported by some of the results of \citet{Bhatia2020}).
While the specific trend motivating that explanation (the model rather uniformly over-predicting \fe\ for F-type stars but not G-type stars) is no longer as clear in the present work, the idea may still warrant exploration.
In Appendix~\ref{appendix:mag-sup} we present a simple, theoretical treatment of convection in the presence of a magnetic field, in which we derive a magnetic suppression factor that allows interpretation of model over-predictions of \fe\ in terms of an ambient magnetic field strength, for which we find values on the order of 10--40~G.
We note that this treatment is very simplified and speculative, and it is only able to treat cases where the model over-predicts \fe.
However, we hope that it inspires further analysis.

An additional path to comparing observed and modeled flicker may be to replace the \fe\ metric with some other quantity or quantities.
A number of statistical regression techniques attempt to fit the variability of stochastic processes with multiple parameters, and this may provide a richer characterization of stellar flicker, which may in turn provide clearer insights on trends in the data.
Examples include ARIMA models \citep[applied to \kep\ data by, e.g.,][]{Feigelson2018,Caceres2019,Caceres2019a}, which fit a set number of parameters describing how each point in the light curve relates to the preceding $n$ points, and Gaussian process regression \citep[applied to \kep\ data by, e.g.][]{Pereira2019}, in which the model is described by a number of specified kernels which are fit to the data, each intended to describe different sources of stochasticity.
Using these sorts of approaches to characterize granular flicker in \kep\ stars, and producing model-predicted values of these metrics as a function of stellar parameters (perhaps by building on the model of \citealp{Samadi2013a,Samadi2013b}) may open a new arena in which granular models and observations can be compared.

\section{Conclusions}
\label{sec:conclusions}

Building on the modeling work of \citet{Samadi2013a,Samadi2013b} and \citet{Cranmer2014}, we have confronted model predictions of \fe\ (the RMS amplitude of $<8$~hr stellar variability) with a larger sample of observational measurements by \citet{Bastien2016}, allowing the discrepancy between observed and modeled \fe\ to be analyzed over a wider range of stellar types.
We have also refined that model.
In every aspect in which the model previously was cast as a theoretical scaling relation relative to a solar value, our model instead draws on scaling relations from grids of numerical simulations covering a range of stellar types, and it directly calculates \fe\ from the other modeled quantities.
This includes Mach numbers calculated in a way that incorporates the stellar metallicity and allows for divergences from ideal theory near the surface of the photosphere through reference to the numerical simulations of \citet{Tremblay2013}, and granular sizes calculated via the scaling relation of \citet{Trampedach2013a} (which removes the free parameter $\beta$ in the original derivation of \citet{Samadi2013a,Samadi2013b}).
We also consider multiple numerical experiments when determining the scaling of the temperature contrast $\Theta\RMS$ with respect to the Mach number, and we attempt to account for the spread in these simulated values through an ``envelope fit'', which produces a range of plausible \fe\ values for a given set of stellar parameters.
Additionally, we have included a term to correct for the influence of \kep 's bandpass on \fe.

These changes to the model have improved its agreement with the observations (reducing the typical misprediction from a factor of 2.5 to a factor of 2), especially when using the envelope fit (which can explain the observed \fe\ of 78\% of our \kep\ sample), though some disagreement remains.
With respect to this disagreement, we have ruled out rotation period and the signatures of large-scale magnetic activity as possible explanations for most convecting dwarf stars, and we have shown that the status of a star as a binary may have an influence on observed \fe, though a larger sample of known binaries is required to confirm this.
We have also discussed other possible influences not considered in detail in the present work.

Given the reliance of the model on results from numerical simulations, the model will be enhanced by further constraints from simulations---particularly improved constraints on the dependence of granular temperature contrast on stellar parameters.
Additionally, it may provide a route for using \fe\ to place constraints on some granular properties for convective simulations.

With interest in exoplanet discovery and characterization as high as ever, stellar flicker \citep[as well as the closely-related radial-velocity jitter, e.g.][]{Hojjatpanah2020,Luhn2020} remains relevant as a source of noise for these observations, and this work has made progress toward fully modeling this noise source from a theoretical perspective.

\acknowledgements

The authors thank Fabienne Bastien and Keivan Stassun for discussions of their past work on the topic, and the anonymous reviewer and the AAS statistics editor for their comments which have strengthened this article.

This work was supported by start-up funds from the Department of Astrophysical and Planetary Sciences at the University of Colorado Boulder, and by the National Science Foundation (NSF) under grant 1613207.
This research has made use of NASA’s Astrophysics Data System Bibliographic Services.

Hinode is a Japanese mission developed and launched by ISAS/JAXA, collaborating with NAOJ as a domestic partner, NASA and STFC (UK) as international partners.  Scientific operation of the Hinode mission is conducted by the Hinode science team organized at ISAS/JAXA.  This team mainly consists of scientists from institutes in the partner countries.  Support for the post-launch operation is provided by JAXA and NAOJ(Japan), STFC (U.K.), NASA, ESA, and NSC (Norway).

This paper draws upon measurements from data collected by the Kepler mission and obtained from the MAST data archive at the Space Telescope Science Institute (STScI). Funding for the Kepler mission is provided by the NASA Science Mission Directorate. STScI is operated by the Association of Universities for Research in Astronomy, Inc., under NASA contract NAS 5–26555.

Guoshoujing Telescope (the Large Sky Area Multi-Object Fiber Spectroscopic Telescope LAMOST) is a National Major Scientific Project built by the Chinese Academy of Sciences. Funding for the project has been provided by the National Development and Reform Commission. LAMOST is operated and managed by the National Astronomical Observatories, Chinese Academy of Sciences.

\software{Astropy version 4.2 \citep{astropy1,astropy2}, Matplotlib version 3.3.3 \citep{matplotlib,matplotlib3.3.3}, NumPy version 1.19.5 \citep{numpy}, SciPy version 1.6.0 \citep{scipy,scipy1.5.3}}

\newpage

\appendix
\section{Derivation of \kep\ Bandpass Correction Term}
\label{appendix:bandpass}

The \kep\ bandpass correction term, described in Section~\ref{sec:bandpass}, accounts for the fact that at different temperatures, different fractions of a stellar spectrum will fall within the \kep\ bandpass, meaning observed variations in flux have a complex dependence on stellar temperature.
The derivation of our correction factor involves a modification of Equations (A.16) through (A.29) of \citet{Samadi2013a}, and here we provide an outline of how our correction modifies key quantities in the original derivation.

We begin by generalizing the bolometric intensity $B = \sigma_{\rm B} T^4/\pi$ to $B \propto T^m$.
This yields
\begin{equation}
	\Delta B = \left(\left(1 + \Theta\right)^m - 1 \right) \left<B\right>_t
\end{equation}
and a Taylor expansion of
\begin{equation}
	\Delta B = \left( m\Theta + \frac{m(m-1)}{2} \Theta^2 \right) \left<B\right>_t.
\end{equation}
This provides an expression for $\left<\Delta B_1 \Delta B_2\right>$, the correlation product of the instantaneous intensity variation of the granules at two points in space and time, of
\begin{equation}
	\left<\Delta B_1 \Delta B_2\right> / \left<B\right>_t = m^2\left< \Theta_1 \Theta_2 \right> \;+\; \frac{m^2(m-1)}{2}\left< \Theta_1 \Theta_2^2 \right> \;+\; \frac{m^2(m-1)}{2} \left< \Theta_1^2 \Theta_2 \right> \;+\; \frac{m^2(m-1)^2}{4}\left< \Theta_1^2 \Theta_2^2 \right>.
\end{equation}
As in the original derivation, the $\left< \Theta_1 \Theta_2^2 \right>$ and $\left< \Theta_1^2 \Theta_2 \right>$ terms are assumed to be zero, and a quasi-normal approximation allows the $\left<\Theta_1^2 \Theta_2^2\right>$ term to be expanded to $2\left<\Theta_1 \Theta_2\right>^2$, yielding
\begin{equation}
	\left<\Delta B_1 \Delta B_2\right> / \left<B\right>_t = m^2\left< \Theta_1 \Theta_2 \right> \;+\; \frac{m^2(m-1)^2}{2}\left< \Theta_1 \Theta_2 \right>^2
\end{equation}
and eventually producing
\begin{equation}
	\mathcal{F}_\tau(\tau,\nu) = \frac{(2\pi)^2\kappa\rho}{R_s^2}\left[ m^2 \left<\widetilde{\Theta_1\Theta_2}\right> + \frac{m^2(m-1)^2}{2} \tilde{\mathcal{B}}_\Theta\right],
\end{equation}
wherein the $\left<\widetilde{\Theta_1\Theta_2}\right>$ term (derived from the $\left<\Theta_1\Theta_2\right>$ term) was found to be negligibly small and taken to be zero by \citet{Samadi2013a}, and so we do likewise.
This final expression for $\mathcal{F}_\tau$ then differs from that in the original derivation by a factor of $m^2(m-1)^2/144$ (which is equal to 1 when $m=4$).
This factor can be carried through Equations (7) and (13) of \citet{Samadi2013a} to find a bandpass correction factor of
\begin{equation}
	\CBP \equiv \frac{\sigma_\text{corrected}}{\sigma_\text{original}} = \sqrt{\frac{m^2(m-1)^2}{144}}.
	\label{eqn:sigma_corrected}
\end{equation}

\begin{figure}[tp]
	\centering
	\includegraphics[width=0.5\linewidth]{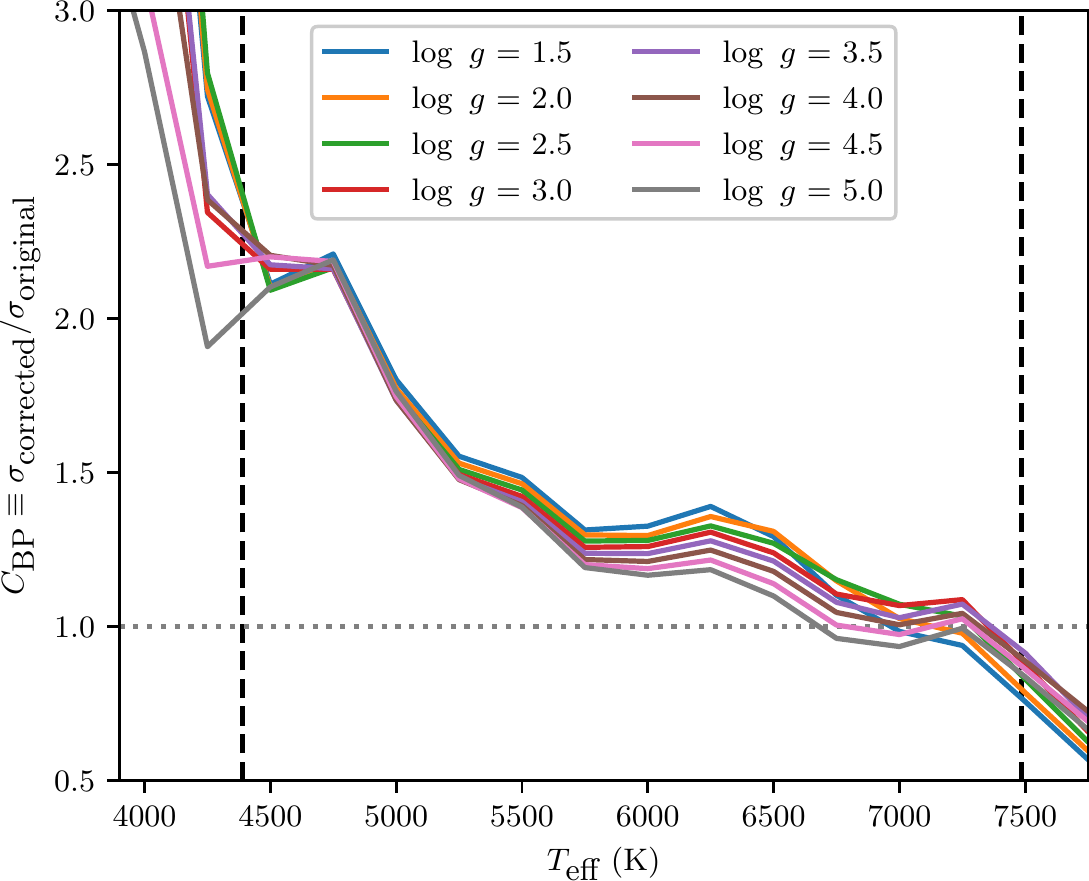}
	\caption{The factor by which our \kep\ bandpass correction factor adjusts the model-predicted $\sigma$ values. Dashed vertical lines indicate the minimum and maximum temperatures within our data catalog.}
	\label{fig:sigma_correction_factor}
\end{figure}

What remains is to determine the values of $m$ to use.
For this task we used the synthetic spectral library of \citet{Lejeune1997}.
For simplicity, we used only their solar-metallicity grid ([M/H] = 0), which contained a collection of 467 stellar spectra with values of \Teff\ between 2,000 and 50,000~K, and \logg\ between $-1$ and 5.5.
For the wavelengths that overlap with the \kep\ passband (i.e., 400--900~nm), the modeled spectra were provided on a grid with a wavelength spacing of 2~nm.
Each spectrum was integrated in two ways: once over all wavelengths to obtain the bolometric flux $F_{\rm bol}$ (which we verified to be equal to $\sigma_{\rm B} \Teff^4$), and once weighted by the
\kep\ spectral response function \citep{Koch2010} to obtain a bandpass-limited flux $F_{\rm Kep}$.
For each subset of models at a fixed value of \logg, we found the logarithmic slope of $F_{\rm Kep} \propto \Teff^m$ by computing
\begin{equation}
	m(T) \, = \, \frac{d \ln F_{\rm Kep}(T)}{d \ln T_{\rm eff}}.
\end{equation}
In concert with Equation~\eqref{eqn:sigma_corrected}, these fitted values produce the scaling factors plotted in Figure~\ref{fig:sigma_correction_factor} and which we apply (interpolating to actual stellar values of (\Teff, \logg)) in Section~\ref{sec:bandpass}.
Our tabulated values, and code using them, are included in our code and data archive \citep{VanKooten2021_Zenodo}.

\section{Calculation of Temperature Contrast from Hinode/SOT data}
\label{appendix:hinode}

\begin{deluxetable}{cccccccccc}
	\tablecaption{\textit{Hinode}/SOT Observations used}
	\tablehead{&&& \colhead{Date} &&& \colhead{Time (UTC)} &&&}
	\startdata
	&&& 2008-01-04 &&& 11:06:26 &&& \\
	&&& 2008-01-05 &&& 00:00:27 &&& \\
	&&& 2008-01-08 &&& 06:13:25 &&& \\
	&&& 2008-02-02 &&& 18:03:55 &&& \\
	&&& 2008-02-03 &&& 18:09:03 &&& \\
	&&& 2008-02-04 &&& 18:27:57 &&& \\
	&&& 2008-02-05 &&& 05:58:57 &&& \\
	&&& 2008-02-05 &&& 10:37:57 &&& \\
	&&& 2008-02-06 &&& 00:02:27 &&& \\
	&&& 2008-02-06 &&& 10:58:27 &&& \\
	\enddata
	\tablecomments{All times given are for the first of the three continuum filters to be imaged. The remaining two images follow within the subsequent 20--40~s.}
	\label{table:hinode-dates}
\end{deluxetable}

To compute a solar value of $\Theta\RMS$, the RMS of $\left( T - \left<T\right>\right) / \left<T\right>$, we turn to the \textit{Solar Optical Telescope} (SOT) on \textit{Hinode} \citep{Kosugi2007,Tsuneta2008}.
Its Broadband Filter Imager (BFI) includes three ``continuum'' filters (in the blue at 450.45~nm, green at 555.05~nm, and red at 668.40~nm, each with a band width of 0.4~nm), and it images with a field of view of 218\arcsec$\times$109\arcsec\ ($\sim$160$\times$80~Mm on the solar surface, containing of order 10,000 granules), a pixel scale of 0.054\arcsec\ ($\sim$40~km on the solar surface), and a diffraction-limited resolution of 0.25\arcsec\ at 500~nm ($\sim$180~km on the solar surface).
We obtained BFI data from the SOT archive for a selection of dates (listed in Table~\ref{table:hinode-dates}), where for each date we used one image from each of the three continuum filters.
Each of the three images were taken in rapid succession ($\sim$10~s cadence) with the same pointing, so that they image very nearly identical conditions (compare the 10~s cadence to typical granular lifetimes of order 10~min).
Each set of images is at or near disk-center, and is a full-resolution, unbinned image (though we note that applying $2\times2$ binning to the level 1 images changes our final results no more than about $5\%$).
Within the range of relevant temperatures, the ratios of fluxes between any two of the three bands maps uniquely to a temperature when assuming blackbody emission, allowing photospheric temperatures to be determined \citep[see, e.g.,][for similar analyses]{Choudhary2013,Watanabe2013,Goodarzi2016}.

Before computing these ratios for any given triplet of images, we first calibrate and align the images.
To calibrate level 1 data units to relative flux, we assume that the mean pixel value in each image corresponds to the mean solar photospheric temperature of 5770~K.
We uniformly scale the pixel values of each image so that the ratios of the means of the three images are equal to the ratios of the fluxes of a 5770~K black body at the three filters' central wavelengths.
We next correct for small, fixed, filter-dependent variations in translation and pixel scale between images \citep[see][]{Shimizu2007}, as well as any small offsets in pointing that may occur between images.
Within each image triplet, we find the translations and scalings that maximize the sum of products of the corresponding pixels for each pair of images.
These values are consistent with those of \citet{Shimizu2007}.
For our final, aligned images we use the average of the fitted scaling factors across all image triplets (calculated per wavelength pair) in light of the fixed nature of the varying plate scales, but we use directly the best-fit translational factors for each image pair to guard against any slight variations in telescope pointing.

With these calibrated and aligned images, we compute flux ratios and map them to temperatures.
We find that the green-to-red ratio produces maps that are significantly more noisy than the blue-to-red and blue-to-green ratios, and so we use only the latter two ratios, and we use in the rest of our analysis the mean of the two temperature maps from those two ratios.

We find that $\Theta\RMS$, computed from individual temperature maps, ranges from 0.044 to 0.049.
Calculated across all maps, $\Theta\RMS=0.0457$ (or $\pm260$~K from the mean in absolute terms).
Our value is slightly higher than some literature values \citep[see][]{Puschmann2005,Baran2015,Gray2018}, which typically note $150-200$~K variations from the mean in photospheric temperatures.
However, we note that \citet{Beck2013} present measurements of temperature fluctuations in solar simulations which are comparable to our values, as well as measurements of observational temperature fluctuations more comparable to the literature values noted previously, and they show that the simulations, when degraded with an appropriate PSF and modeled stray-light contribution, closely match the observational values.
They interpret this as indicating that the higher temperature fluctuations in the simulations can be reasonably taken as representative of the actual Sun, while the typically-lower observational values are due to spatial and spectral degradation and stray light.
This suggests that our higher observational value may also be reasonable, especially since our use of un-binned BFI images affords us higher spatial resolution than the SP data used by \citet{Beck2013}, and our broadband technique limits the effect of spectral degradation.

\section{Avenues Toward Understanding Magnetic Suppression of Flicker}
\label{appendix:mag-sup}

Here we present a simple discussion of small-scale magnetic fields in the presence of convection and speculate on how those fields could affect flicker measurements.
Numerous, more advanced approaches exist in the literature \citep[e.g.,][and references therein]{Gough1966, Knoelker1988,Cattaneo2003,MacDonald2014}; however, our discussion is based on first-principles arguments of convective force balances, which have very successfully described convection in the presence of both strong and weak global rotation \citep[reviewed and applied to the Sun respectively in][]{aurnou_etal_2020, vasil_etal_2020}.

We assume an ideal gas with an equation of state of $p = R \rho T$ and write the fully compressible, magnetohydrodynamic (MHD) momentum equation,
\begin{equation}
	\partial_t \bm{u} + \bm{u}\dot\grad\bm{u} + \frac{1}{\rho}\bm{J}\times\bm{B} = -\frac{1}{\rho}\grad p + \bm{g},
\end{equation}
where we assume ideal MHD so that $\bm{J} = (\grad\times\bm{B})/4\pi$.
We next decompose thermodynamics into background (subscript 0) and fluctuating (subscript 1) pieces, and we assume that the fluctuations are relatively small compared to the background (which may not be a good assumption for all of the stars in this work).
We assume that the background pressure gradient is in hydrostatic equilibrium, and subtract these background terms from the above equation to find
\begin{equation}
	\partial_t \bm{u} + \bm{u}\dot\grad\bm{u} + \frac{1}{\rho_0}\bm{J}\times\bm{B} = -\frac{1}{\rho_0}\grad p_1,
	\label{eqn:momentum_fluctuations}
\end{equation}
where we have assumed that $\rho^{-1} \approx \rho_0^{-1}$ for small thermodynamic perturbations.

In the absence of magnetism and in a statistically stationary state, the advective forces should roughly balance the pressure gradient,
\begin{equation}
	\bm{u}\dot\grad\bm{u} \sim -\frac{1}{\rho_0}\grad p_1.
\end{equation}
Expressing the gradient as an inverse characteristic length scale ($\grad \approx L^{-1}$) and dividing this balance by the square sound speed ($c_s^2 = \partial p/\partial \rho = R T_0$), we find that
\begin{equation}
	\frac{|u|^2}{L c_s^2} \sim \frac{p_1}{L p_0}
	\qquad\Rightarrow\qquad
	\mathcal{M}^2 \sim \frac{T_1}{T_0}.
	\label{eqn:hydro_balance}
\end{equation}
The latter expression relies on the linearized equation of state \citep[per, e.g., Equation~(11) of][]{brown_etal_2012} to assume that $p_1/p_0 \sim T_1/T_0$, and retrieves the classical expression that the squared Mach number scales like the temperature perturbations \citep[discused in Section~\ref{sec:theta} and verified in][]{anders_brown_2017}.

Most careful treatments of magnetoconvection have focused on the case where there is a strong background magnetic field \citep[for a brief review, see][]{plumley_julien_2019}.
However, to understand stellar magnetism outside of starspots we are interested in the less-understood case where there is no defined, strong, background magnetic field and fluctuations dominate.
We now \emph{speculate} that this magnetoconvection at the stellar surface exhibits a triple force balance between nonlinear magnetic forces, nonlinear inertial forces, and the pressure gradient \citep[inspired by the Coriolis-inertial-Archimedean (or CIA) balance of rapidly rotating convection, see Equation~(24) of][]{aurnou_etal_2020}.
Put differently, we assume that induction generates magnetic fields whose amplitudes saturate once they are strong enough to feed back on the convection which generates them.
Assuming this balance and following the same arguments as the non-magnetized balance, we retrieve
\begin{equation}
	\frac{1}{\rho_0}\bm{J}\times\bm{B} \sim \bm{u}\dot\grad\bm{u} \sim -\frac{1}{\rho_0}\grad p_1
	\qquad\Rightarrow\qquad
	\frac{|B^2|}{4\pi \rho_0 c_s^2} \sim \mathcal{M}^2 \sim \frac{T_1}{T_0}.
	\label{eqn:B_triple_balance}
\end{equation}
This balance allows for an immediate estimation of the magnitude of small-scale magnetic fields at the stellar surface in cgs units,
\begin{equation}
	|B| \sim \sqrt{4\pi\rho_0} c_s \mathcal{M}.
	\label{eqn:B_est}
\end{equation}
At the Sun's surface, $\rho_0 \approx 2 \times 10^{-7}$ g/cm$^{-3}$ and $c_s \approx 10^6$ cm s$^{-1}$ \citep{avrett_loeser_2008}, for an approximate magnitude of $|B| \sim 10^{3} \mathcal{M}$, which suggests magnetic field strengths of order 100~G for Mach numbers of order 0.1 at the solar surface.
This is quite a reasonable estimate; for one of many observational estimates of typical, quiet-sun magnetic field strengths, see \citet{OrozcoSuarez2012}, who find a distribution peaking near 100~G.
This suggests that the triple-balance of Equation~\eqref{eqn:B_triple_balance} is a plausible starting point.

Armed with this assumption, we return to Equation~\eqref{eqn:momentum_fluctuations} and take a time- and volume- average (represented as an overbar).
Continuing to assume a statistically-stationary flow, the mean forces must satisfy
\begin{equation}
	\overline{\bm{u}\dot\grad\bm{u}} = -\overline{\frac{1}{\rho_0}\left(\grad p_1 + \bm{J}\times\bm{B} \right)} \approx -f\overline{\frac{1}{\rho_0}\grad p_1}.
	\label{eqn:average_forces}
\end{equation}
Rearranging Equation~\eqref{eqn:average_forces} and applying the procedure used to derive Equation~\eqref{eqn:hydro_balance}, we find
\begin{equation}
\frac{T_1}{T_0} \sim \frac{\mathcal{M}^2}{f}.
\label{eqn:mag_suppression}
\end{equation}
Here we have assumed that the Lorentz force and pressure gradient have the same magnitude, but we have left $f$ as a free parameter which describes how these vectors are on average oriented with respect to one another ($f = 2$ for uniformly parallel vectors and $f = 0$ for antiparallel).
A value of $f > 1$ corresponds to magnetic suppression of convection in Equation~\eqref{eqn:mag_suppression} (i.e. a reduction in the effective \Ma\ that is controlling the temperature fluctuations); it has been known for decades that strong magnetic fields suppress convection \citep{chandrasekhar_1961}, and it would make sense for weak fields to have a similar (but smaller) effect.
If the influence of magnetism as considered here is the cause for our remaining model--observation discrepancy (or some portion thereof), values of $f$ greater than one are the values that will reduce that remaining discrepancy.
That discrepancy is most typically a model overprediction by a factor of about 2 (see Section \ref{sec:initial-comp}), and since $\fe \propto \left(T_1 / T_0 \right)^2$, this misprediction would be resolved by a factor of $f \sim \sqrt{2}$ (neglecting for now any differences in the amplitudes of the forces in the triple balance).

A determination of the validity of this treatment is beyond the scope of this work.
We leave the reader with a few questions which could be answered by targeted magnetohydrodynamic simulations of stellar surface convection:
\begin{enumerate}
	\item Is the triple-balance described in Equation~\eqref{eqn:B_triple_balance} seen in evolved simulations of nonlinear MHD convection?
	If so, Equation~\eqref{eqn:B_est} provides a straightforward way of estimating the magnitude of surface magnetism for stars with convective envelopes.
	\item Regardless of whether the triple-balance is achieved, is the assumption that introduces the magnetic suppression factor $f$ in Equation~\eqref{eqn:average_forces} valid?
	If so, what is the magnitude of $f$?
\end{enumerate}


\end{document}